\documentclass[10pt]{article} % For LaTeX2e
\usepackage[accepted]{tmlr}

\usepackage{amsmath}
\usepackage{amssymb}
\usepackage{amsfonts}
\usepackage{amsmath,amsfonts,bm}

\def\eqref#1{equation~\ref{#1}}
\def\1{\bm{1}}

\DeclareMathAlphabet{\mathsfit}{\encodingdefault}{\sfdefault}{m}{sl}
\SetMathAlphabet{\mathsfit}{bold}{\encodingdefault}{\sfdefault}{bx}{n}

\usepackage{microtype}
\usepackage[inline]{enumitem}
\usepackage{graphicx}
\usepackage{subcaption}
\usepackage[T1]{fontenc}
\usepackage{wrapfig}
\usepackage{booktabs}
\usepackage{tabularx}
\usepackage{makecell}
\usepackage{multirow}
\usepackage{longtable}
\usepackage{colortbl}
\usepackage{nicefrac}
\usepackage{xcolor}
\usepackage[most]{tcolorbox}
\usepackage{array}         % for column formatting
\usepackage{booktabs}      % for better rules
\tcbuselibrary{listings,skins,breakable}

\usepackage{xfp}
\usepackage{url}
\usepackage{hyperref}

\title{Terminal Agents Suffice for Enterprise Automation}

\newif\ifrevhighlight\revhighlightfalse

\def\month{10}  % Insert correct month for camera-ready version
\def\year{2026} % Insert correct year for camera-ready version
\def\openreview{\url{https://openreview.net/forum?id=6zrEOjtoVF}} % Insert correct link to OpenReview for camera-ready version

\author{\name Patrice Bechard \email patrice.bechard@servicenow.com \\
      \addr ServiceNow
      \AND
      \name Orlando Marquez Ayala \email orlando.marquez@servicenow.com \\
      \addr ServiceNow
      \AND
      \name Emily Chen \email emily.chen@servicenow.com\\
      \addr ServiceNow
      \AND
      \name Jordan Skelton \email jordan.skelton@servicenow.com\\
      \addr ServiceNow
      \AND
      \name Sagar Davasam \email sagar.davasam@servicenow.com\\
      \addr ServiceNow
      \AND
      \name Srinivas Sunkara \email srinivas.sunkara@servicenow.com\\
      \addr ServiceNow
      \AND
      \name Vikas Yadav \email vikas.yadav@servicenow.com\\
      \addr ServiceNow
      \AND
      \name Sai Rajeswar \email sai.mudumba@servicenow.com\\
      \addr ServiceNow, Mila -- Quebec AI Institute, Universit\'e de Montr\'eal
}

\begin{document}

\maketitle

\begin{abstract}
There has been growing interest in building agents that can interact with digital platforms to execute meaningful enterprise tasks autonomously. Among the approaches explored are tool-augmented agents built on abstractions such as Model Context Protocol (MCP) and web agents that operate through graphical interfaces. Yet, it remains unclear whether such complex agentic systems are necessary given their cost and operational overhead. We argue that a coding agent equipped only with a terminal and a filesystem can solve many enterprise tasks more effectively by interacting directly with platform APIs. We evaluate this hypothesis across diverse real-world systems and show that these low-level terminal agents match or outperform more complex agent architectures at a fraction of the cost. Our findings suggest that simple, flexible programmatic interfaces combined with strong foundation models should be the backbone of enterprise automation.
\end{abstract}

\section{Introduction}

\begin{figure}[ht]
\begin{center}
\includegraphics[width=\textwidth]{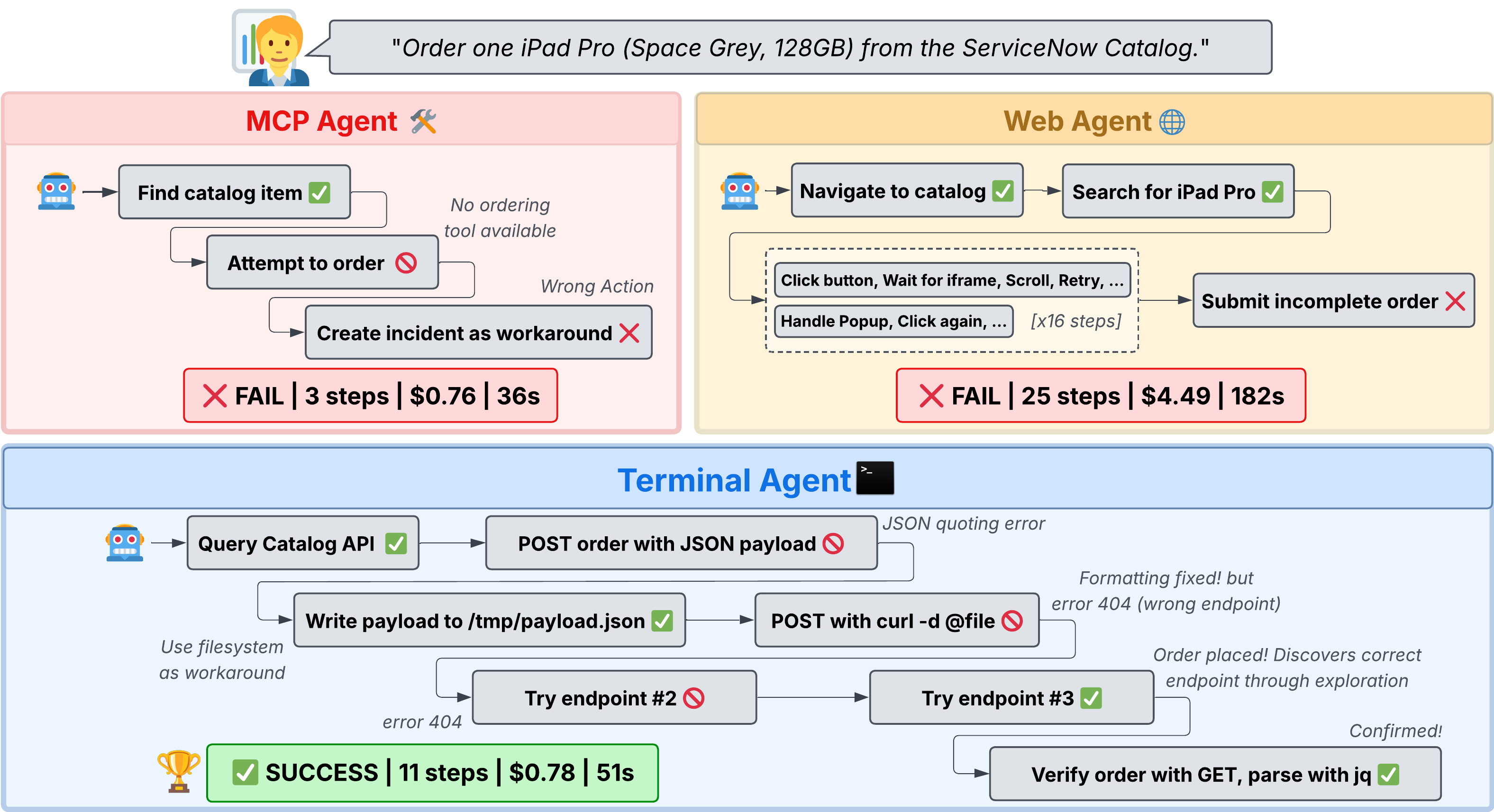}
\end{center}
\vspace{-5pt}
\caption{\small{\textbf{Execution traces of agents ordering an iPad Pro.} (Top-left) The MCP agent identifies the catalog item but cannot proceed without an ordering tool. It falls back to creating a support ticket and fails. (Top-right) The web agent reaches the catalog page but becomes confused within the iframe-based UI, leading to a long, costly trajectory that fails to complete the order. (Bottom) The terminal agent encounters JSON quoting errors when constructing the request payload and 404 responses from incorrect API endpoints, but recovers by writing the payload to a temporary file and exploring alternative endpoints. It completes the task at an order of magnitude lower cost than the web agent, demonstrating the flexibility, resilience, and efficiency of direct terminal-based interaction.}}
\vspace{-5pt}
\end{figure}

Large Language Models (LLMs) have rapidly evolved from code completion assistants~\citep{austin2021programsynthesislargelanguage, chen2021evaluatinglargelanguagemodels} to agents capable of executing multi-step tasks across software systems~\citep{jimenez2024swebench, zhou2024webarena, mcpuniverse}. In enterprise settings, this shift is particularly consequential. Rather than drafting text or answering questions, LLM-powered agents are increasingly expected to perceive system state, reason about business context, and perform actions that modify operational data~\citep{huang-etal-2025-crmarena, boisvert2024workarena, malay2026enterpriseopsgymenvironmentsevaluationsstateful}. These agents operate over long horizons and interact with production systems where errors can cascade across records, approvals, and downstream workflows~\citep{gupta2026world, nair2026enterprise}. Enterprise automation therefore introduces new challenges, requiring agents to reliably interact with complex platforms and operate under real-world constraints.

To address these challenges, two architectural directions have become especially prominent. Web agents operate through a graphical user interface (GUI), issuing low-level actions over DOM elements and screenshots~\citep{deng2023mindweb, qin2025ui}. Tool-augmented agents~\citep{yao2022react, qin2023toolllm, patil2025the} expose curated action schemas through frameworks like Model Context Protocol (MCP)~\citep{anthropic2024mcp}, enabling models to invoke predefined operations. Although these approaches differ in implementation, they share a common design choice: introducing structured abstractions between the model and the underlying platform, rather than enabling direct interaction.

Abstractions, however, come with tradeoffs. GUI agents must reason over long, brittle action chains that are sensitive to interface changes~\citep{prabhu2026walt}, while curated tool registries simplify invocation at the cost of restricting expressivity to predefined operations and compositions~\citep{wang2024executable}. Recent generalist code agents such as Claude Code~\citep{anthropic_claude_code_2026} and OpenClaw~\citep{steinberger_openclaw_2026} demonstrate that strong performance on complex, real-world tasks can emerge without heavy, pre-curated abstraction layers by operating directly over programmable interfaces such as command-line environments and application programming interfaces (APIs). This follows the spirit of the Bitter Lesson~\citep{sutton2019bitterlesson}: general methods that let strong models leverage flexible, computation-friendly interfaces tend to outperform approaches built on extensive hand-engineered structure, and in this work we test whether the same holds for enterprise agent interfaces specifically. Modern enterprise platforms already expose similarly expressive APIs for querying data, updating records, and performing complex operations programmatically; in such settings, additional wrappers may reduce flexibility rather than enhance it, trading generality for structure without clear necessity.

In this work, we empirically test whether additional abstraction layers are necessary when stable APIs are available. We construct realistic enterprise automation benchmarks across multiple production-grade platforms and compare three paradigms: GUI-driven agents, tool-augmented agents, and minimal terminal-based coding agents that operate through direct API interaction. Across diverse workflows, \textbf{minimal terminal agents match or outperform more complex architectures despite their simplicity, while maintaining competitive efficiency}. This positions the coding agent as the foundation rather than one option among three: a terminal and filesystem are the substrate enterprise automation should build from, extended with persistent skills or browser access when a task requires it. These findings challenge the prevailing assumption that increasingly sophisticated agent stacks are required for enterprise automation, suggesting instead that strong foundation models combined with direct programmatic interfaces may suffice for a broad class of real-world tasks. Our main contributions are as follows:

\begin{itemize}

\item We show that simple terminal agents operating through direct API interaction are both effective and efficient for enterprise automation, outperforming MCP-based tool-augmented agents and matching or exceeding web-agent performance at substantially lower cost, based on a systematic evaluation of frontier LLMs across these three paradigms on diverse real-world enterprise tasks.

\item We introduce a unified benchmark spanning multiple production platforms, including verified evaluation environments and datasets capturing realistic enterprise tasks.

\item We further study practical extensions to terminal agents, including self-created reusable skills and browser access as a fallback for genuinely UI-bound tasks.

\end{itemize}

Resources supporting this work, including evaluation environments, datasets, and code, are publicly available through the StarShell GitHub repository (\url{https://github.com/ServiceNow/StarShell}) and Hugging Face collection (\url{https://huggingface.co/collections/ServiceNow-AI/starshell}), facilitating reproducibility and further research on enterprise agent systems.

\section{Related Work}
\textbf{Web and GUI agents in enterprise benchmarks.}
A growing body of work studies agents that complete tasks through web or GUI interaction, translating natural-language instructions into sequences of clicks, typing, and navigation~\citep{deng2023mindweb, zhou2024webarena}. WebArena~\citep{zhou2024webarena} introduced a realistic and reproducible web environment and showed that even strong agents remain far below human performance on end-to-end success, particularly on tasks requiring long interaction chains. In enterprise settings, WorkArena~\citep{drouin2024workarena} focuses on knowledge-work tasks on ServiceNow, and follow-up work~\citep{boisvert2024workarena} extends this setting to hundreds of compositional workflows, revealing planning, reasoning, and retrieval as key bottlenecks for current web-agent designs. Complementary benchmarks explore other enterprise platforms, including SCUBA for Salesforce CRM workflows~\citep{dai2025scuba} and TheAgentCompany for agents operating across multiple enterprise tools and services~\citep{xu2025theagentcompany}.

\textbf{API-first and coding-agent approaches.}
A parallel line of work explores programmatic interaction as an alternative to GUI control, arguing that APIs and executable code provide more reliable and compact interfaces than step-by-step UI navigation. Beyond Browsing~\citep{song-etal-2025-beyond} evaluates this idea on WebArena by augmenting agents with API access, showing that API-enabled agents outperform browsing-only agents, while hybrid agents combining APIs and browsing perform best. AXIS~\citep{lu-etal-2025-axis} reaches similar conclusions in desktop environments, demonstrating that API-first ``skills'' can significantly reduce task completion time. Related work on coding agents treats executable code as the primary interface to the environment. CodeAct~\citep{wang2024executable} frames agent actions as executable code, enabling direct interaction with external systems. AppWorld~\citep{trivedi2024appworld} introduces interactive coding environments with programmatic evaluation that verifies both intended state changes and unintended side effects. Gorilla~\citep{patil2023gorilla} shows that grounding models in retrieved API documentation improves reliability when interacting with evolving APIs. Our work adopts the same philosophy: rather than encoding task-specific structure into the interface itself, we expose a general, flexible interaction surface and let a sufficiently capable model discover and compose the operations it needs.

\textbf{Externalizing context through environment interaction.}
Enterprise automation tasks require agents to operate over information that exceeds the limits of a single prompt, motivating approaches that externalize context into the environment. Recursive Language Models (RLMs)~\citep{zhang2025recursive} frame long-context processing as interaction with external tools through a REPL-style loop, while software-oriented agents such as SWE-agent~\citep{yang2024swe} and OpenHands~\citep{wang2024openhands} expose the filesystem and execution environment as part of the agent state. In parallel, work on memory and experience accumulation explores how agents can persist and reuse information across tasks~\citep{xu2025amem, ouyang2025reasoningbank, nekoei2025just, zhang2025agentic}. Our work adopts this perspective, using the filesystem and APIs as a simple mechanism for managing context and enabling reusable skills in enterprise automation tasks.

\vspace{-3pt}
\section{StarShell}
\vspace{-3pt}

We investigate whether terminal-based coding agents can serve as the foundation for enterprise automation when APIs are available. To test this hypothesis, we compare three agent paradigms that differ only in their interaction modality: (1) GUI-driven web agents operating through browser interfaces, (2) tool-augmented agents accessing curated APIs via Model Context Protocol (MCP), and (3) minimal terminal agents interacting directly with platform APIs. All agents use the same LLM backbone and are evaluated on identical enterprise benchmarks to isolate the effect of the interaction paradigm.

\subsection{Agent Interaction Paradigms}

\textbf{Tool-augmented agents} operate through a curated set of API tools exposed via MCP servers, corresponding to operations such as listing records, creating entries, or updating fields. We refer to these interchangeably as \emph{tool-use agents}, the term used when reporting results. The agent selects a tool and provides arguments for invocation. This abstraction simplifies execution by compressing complex interactions into high-level operations, but constrains the agent to the functionality exposed by the tool registry.

\textbf{Web agents} operate through graphical interfaces, observing the rendered UI and issuing low-level actions such as clicking, typing, or navigating pages. In our implementation, web agents operate through the official Playwright MCP server~\footnote{\url{https://github.com/microsoft/playwright-mcp}, v0.0.68.}, which exposes 21 tools spanning accessibility-tree snapshots, screenshots, form and keyboard interaction, tab management, and JavaScript evaluation (Appendix~\ref{sec:playwright-mcp-server}). Agents are prompted to inspect page structure before interacting with elements and to verify the resulting state afterwards (Appendix~\ref{sec:prompts}).

Our primary focus is a \textbf{simple coding agent that operates through a terminal and filesystem}. Instead of invoking predefined tools, it writes and executes code to interact directly with platform APIs, enabling flexible API interaction, data manipulation, and exploration. This pattern mirrors modern coding agents such as Claude Code and OpenClaw, where the model operates in a REPL-style loop of reasoning, execution, and environment inspection.

\begin{figure}[t]
\begin{center}
\includegraphics[width=0.9\textwidth]{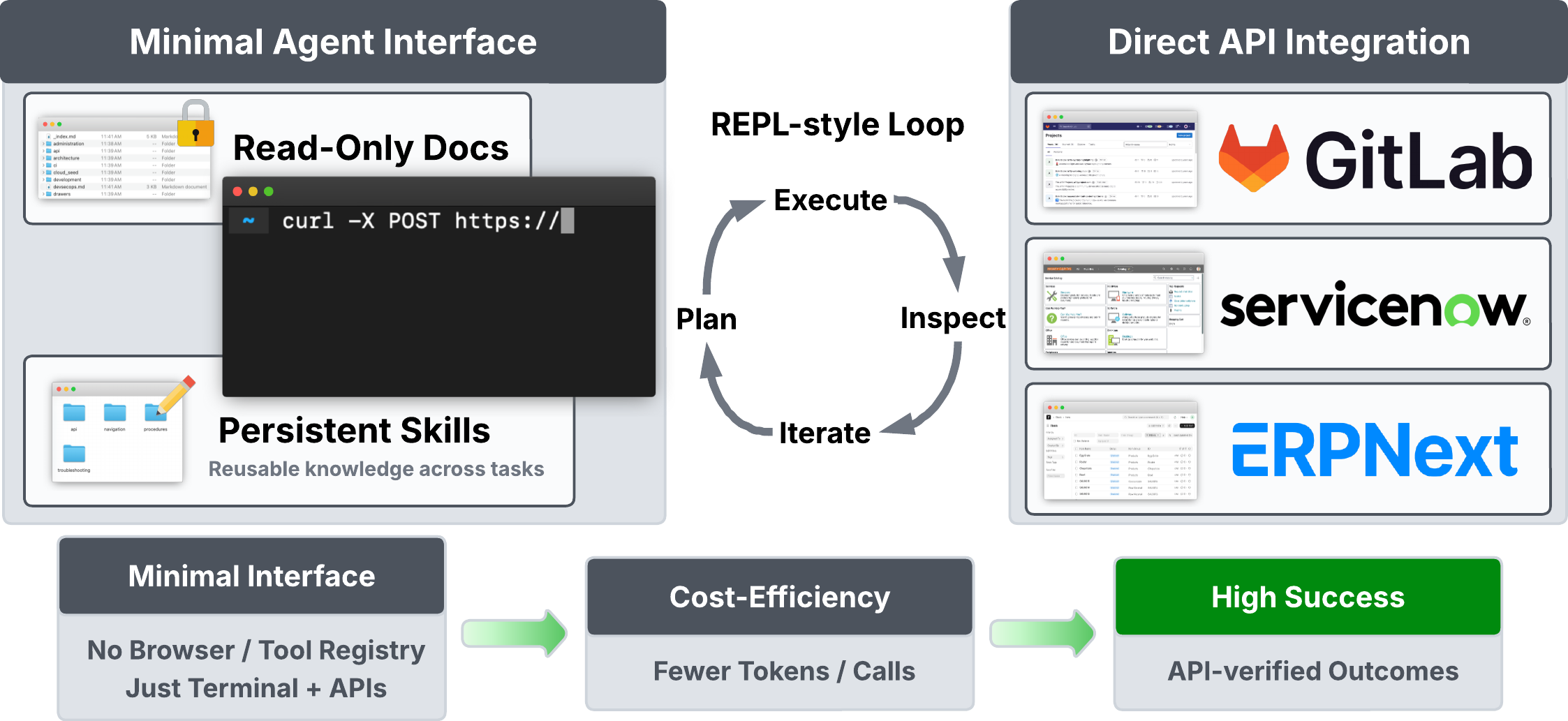}
\end{center}
\vspace{-7pt}
\caption{\small{\textbf{Overview of StarShell: a minimal terminal agent for enterprise automation.} The agent operates through a terminal and filesystem, optionally using documentation and persistent skills to discover and invoke APIs directly on enterprise platforms (e.g., GitLab, ServiceNow, ERPNext), without relying on GUI interaction or pre-defined tool registries.}}
\vspace{-5pt}
\end{figure}

\subsection{StarShell: A Terminal-Based Enterprise Agent}

We implement this terminal agent as StarShell, a minimal coding-agent environment for enterprise automation tasks, operating through two primary interfaces: a terminal for executing commands and a filesystem for storing artifacts. For each task, the agent receives the task description and execution context, then iteratively generates commands or code snippets to run. Typical actions include querying platform APIs, filtering results, updating records, or generating scripts to automate repetitive operations. API responses and execution outputs are returned to the model as observations, enabling iterative reasoning and correction. The filesystem provides persistent task state: the agent can read documentation, cache intermediate results, and persist reusable ``skills'' such as scripts or notes capturing previously discovered solutions.

Unlike tool-based agents, StarShell does not rely on predefined action schemas. Instead, it discovers platform capabilities dynamically by reading documentation or inspecting API responses. This design allows the agent to compose operations that may not be represented in curated tool registries.

\subsection{Enterprise Benchmark Environments}

We evaluate agents across three enterprise platforms: ServiceNow,\footnote{Docs: \url{https://www.servicenow.com/docs/}; MCP: \url{https://github.com/echelon-ai-labs/servicenow-mcp}, commit \texttt{0625060}.} GitLab,\footnote{Docs: \url{https://docs.gitlab.com/}; MCP: \url{https://github.com/zereight/gitlab-mcp}, v2.0.30.} and ERPNext,\footnote{Docs: \url{https://docs.erpnext.com/}; MCP: \url{https://github.com/rakeshgangwar/erpnext-mcp-server}, commit \texttt{e20278b}.} which represent common categories of enterprise software: IT service management, software development lifecycle management, and enterprise resource planning. Table~\ref{tab:benchmark-summary} summarizes the benchmark statistics.

\begin{wraptable}{r}{0.48\textwidth}
\vspace{-8pt}
\centering
\small
\setlength{\tabcolsep}{4pt}
\begin{tabular}{lccc}
\toprule
& \textbf{ServiceNow} & \textbf{GitLab} & \textbf{ERPNext}\\
\midrule
Samples        & 330 & 192 & 207 \\
MCP Tools      & 83 & 81 & 7 \\
Doc. Pages     & 61k & 2.65k & 5.41k \\
\bottomrule
\end{tabular}
\vspace{-5pt}
\caption{\small{Summary statistics of the evaluation benchmark across three enterprise platforms.}}
\label{tab:benchmark-summary}
\vspace{-12pt}
\end{wraptable}

Each benchmark consists of natural-language tasks that require agents to inspect system state, retrieve information, and perform actions that modify platform records. The tasks range from simple record queries to multi-step workflows involving filtering, conditional updates, and reasoning across multiple objects. An exhaustive list of task types for each environment is provided in Appendix~\ref{sec:task-types}.

For ServiceNow and GitLab, we build on tasks introduced in prior work~\citep{drouin2024workarena, zhou2024webarena}, but adapt the evaluation pipelines to use programmatic verification against the live platform rather than hardcoded values or browser-script checks. For ERPNext, we construct a new benchmark of enterprise workflows spanning multiple record types and task complexities. Details are provided in Appendix~\ref{sec:task-creation}.

All environments are deployed in containerized instances. We reuse the GitLab container setup from~\citet{hattami2025webarena} and build custom environments for ServiceNow and ERPNext. For each platform, we provide (1) a sandbox environment with isolated data, (2) API access for programmatic interaction, (3) optional MCP tool registries exposing curated platform operations, and (4) a local documentation corpus accessible through the filesystem. Documentation is obtained by scraping official docs or repository sources and converted to markdown for standardized access. We selected widely used open-source MCP server implementations based on public availability, platform coverage, and community adoption. We do not customize or expand these registries, as our goal is to evaluate practical MCP-based agents in realistic off-the-shelf conditions rather than through platform-specific tool engineering. Throughout, tool counts are the number of tools each server actually advertises to the model in the configuration we run, which can be smaller than the number a server implements: some servers gate groups of tools behind opt-in flags that we leave at their defaults, and some list tools in configuration files that ship without an implementation. We report what the agent could invoke rather than what the repository documents.

\vspace{-5pt}
\subsection{Experimental Setup}

\textbf{Agent implementations.} All agents are implemented using the OpenAI Agents SDK~\citep{openai_agents_python}. Web agents use a Playwright MCP server that exposes browser actions and observations. Tool-augmented agents access curated platform APIs via MCP servers for ServiceNow, GitLab, and ERPNext. Terminal agents operate in a sandboxed environment with terminal and filesystem access. The tools exposed by the platform-specific MCP servers are listed in Appendix~\ref{sec:mcp-servers}, those of the Playwright MCP server in Appendix~\ref{sec:playwright-mcp-server}.

We evaluate multiple frontier LLM backbones, including Claude Sonnet 4.6~\citep{anthropic2026claudesonnet46}, Claude Opus 4.6~\citep{anthropic2026claudeopus46}, GPT-5.4 Thinking~\citep{openai2026gpt54systemcard}, and Gemini 3.1 Pro~\citep{googledeepmind2026gemini31pro}. Within each comparison, all agent paradigms share the same backbone to ensure fairness.

\textbf{Experimental design.} Our experiments proceed in two stages. We first compare the three interaction paradigms (web agents, tool-augmented agents, and terminal agents) under a minimal configuration without documentation access or reusable skills. This isolates the effect of the interaction interface itself. We then introduce additional capability modules, such as documentation access and skill persistence, and measure their contribution through controlled ablations.

For each task, the agent receives a natural-language description and interacts with the platform until the task is completed or a predefined execution limit is reached. Tasks run in isolated environments to avoid cross-task interference.

\textbf{Metrics.} Our primary metric is \textbf{success rate (SR)}, the percentage of tasks completed according to task-specific verifications over the resulting system state. Because task outcomes are binary, we estimate standard errors for SR using the sample-proportion estimator, and in our result tables we highlight scores within one standard error of the best-performing method.

To evaluate efficiency, we report \textbf{inference cost}, computed from the token usage of the underlying language model. Cost provides a consistent and model-agnostic measure of efficiency that is less sensitive to environment-specific latency than wall-clock time.

Appendix~\ref{sec:full-results} also reports tool calls and wall-clock time; the former ignores reasoning cost and the latter depends on infrastructure latency, so we use cost as our primary efficiency metric.

\section{Experiments}

\begin{table}[t]
\centering
% \small
\setlength{\tabcolsep}{5pt}
\begin{tabular}{l c cc c cc c cc c cc}
\toprule
&&\multicolumn{2}{c}{\textbf{ServiceNow}} &&
  \multicolumn{2}{c}{\textbf{GitLab}} &&
  \multicolumn{2}{c}{\textbf{ERPNext}} &&
  \multicolumn{2}{c}{\textbf{Overall}} \\
\textbf{Agent} && \multicolumn{2}{c}{(330)} &&
  \multicolumn{2}{c}{(192)} &&
  \multicolumn{2}{c}{(207)} &&
  \multicolumn{2}{c}{(729)} \\
\cmidrule(lr){3-4}
\cmidrule(lr){6-7}
\cmidrule(lr){9-10}
\cmidrule(lr){12-13}
&& \textit{SR~(\%)} & \textit{Cost~(\$)} &&
  \textit{SR~(\%)} & \textit{Cost~(\$)} &&
  \textit{SR~(\%)} & \textit{Cost~(\$)} &&
  \textit{SR~(\%)} & \textit{Cost~(\$)} \\
\midrule
\rowcolor{gray!10}\multicolumn{13}{c}{\textit{Claude Sonnet 4.6}} \\
Tool-use            &&
    11.5 & \textbf{0.76} &&
    45.2 & 0.48 &&
    55.6 & \textbf{0.14} &&
    32.9 & \textbf{0.51} \\
Web            &&
    \textbf{72.4} & 4.49 &&
    \textbf{82.9} & 0.88 &&
    61.8 & 3.63 &&
    \textbf{72.2} & 3.29 \\
Terminal            &&
    \textbf{73.6} & 0.78 &&
    76.5 & \textbf{0.28} &&
    \textbf{67.6} & 0.46 &&
    \textbf{72.7} & 0.56 \\
\midrule
\rowcolor{gray!10}\multicolumn{13}{c}{\textit{Claude Opus 4.6}} \\
Tool-use             &&
    16.1 & \textbf{0.66} &&
    46.8 & 0.90 &&
    68.9 & \textbf{0.17} &&
    39.2 & \textbf{0.58} \\
Web            &&
    \textbf{77.6} & 4.21 &&
    \textbf{81.9} & 0.85 &&
    \textbf{81.6} & 6.49 &&
    \textbf{79.9} & 3.97 \\
Terminal            &&
    \textbf{79.1} & 1.94 &&
    \textbf{80.2} & \textbf{0.50} &&
    76.8 & 0.72 &&
    \textbf{78.7} & 1.22 \\
% \midrule
% \rowcolor{gray!10}\multicolumn{13}{c}{\textit{GPT-5.2 (Medium)}} \\
% Tool-use             &&
%     16.7 & \textbf{0.10} &&
%     47.9 & 0.25 &&
%     68.6 & \textbf{0.19} &&
%     39.7 & \textbf{0.16} \\
% Web            &&
%     \textbf{75.8} & 0.66 &&
%     \textbf{72.9} & \textbf{0.18} &&
%     \textbf{76.3} & 1.33 &&
%     \textbf{75.2} & 0.73 \\
% Terminal            &&
%     \textbf{77.3} & 0.22 &&
%     \textbf{71.8} & \textbf{0.17} &&
%     70.5 & 0.48 &&
%     \textbf{73.9} & 0.28 \\
\midrule
\rowcolor{gray!10}\multicolumn{13}{c}{\textit{GPT-5.4 Thinking (Medium)}} \\
Tool-use             &&
    18.5 & \textbf{0.14} &&
    47.9 & 0.40 &&
    62.8 & \textbf{0.21} &&
    38.8 & 0.23 \\
Web            &&
    69.4 & 0.54 &&
    \textbf{81.4} & 0.17 &&
    \textbf{72.5} & 0.51 &&
    \textbf{73.4} & 0.43 \\
Terminal            &&
    \textbf{77.0} & 0.20 &&
    71.3 & \textbf{0.13} &&
    \textbf{70.0} & 0.24 &&
    \textbf{73.5} & \textbf{0.19} \\
\midrule
\rowcolor{gray!10}\multicolumn{13}{c}{\textit{Gemini 3.1 Pro}} \\
Tool-use             &&
    14.2 & \textbf{0.10} &&
    48.9 & 0.15 &&
    62.8 & \textbf{0.07} &&
    37.1 & 0.11 \\
Web            &&
    62.1 & 0.68 &&
    \textbf{84.6} & 0.22 &&
    65.2 & 1.13 &&
    68.9 & 0.69 \\
Terminal            &&
    \textbf{78.5} & \textbf{0.10} &&
    79.8 & \textbf{0.06} &&
    \textbf{73.9} & 0.10 &&
    \textbf{77.5} & \textbf{0.09} \\
\bottomrule
\end{tabular}
\vspace{-5pt}
\caption{\small{\textbf{Main results across agent interaction paradigms}. Success rate (SR, $\uparrow$) and average cost per task ($\downarrow$) for tool-use, web, and terminal agents on three enterprise platforms with four
  backbone LLMs. Bold indicates the best SR and lowest cost within each platform--model group, along with any score within one standard error of the best.}}
\label{tab:results-main}
\vspace{-12pt}
\end{table}

\subsection{Comparing types of agents}\label{sec:agent_types}
\vspace{-5pt}

Table~\ref{tab:results-main} presents results comparing different agent types on tasks from three enterprise environments using 4 different base LLMs.

\textbf{Tool-use agents are limited by their curated tool catalogs.} Across all models, tool-use agents achieve the lowest success rates, particularly on ServiceNow (11.5--18.5\%), where, despite access to 83 tools, many tasks require actions beyond what the available endpoints support. GitLab's catalog is almost exactly the same size at 81 tools, yet performance there is far higher at 45--49\%, though still well below the other agent types; catalog size alone therefore does not determine how far a tool-use agent gets. Paradoxically, the platform where tool-use agents perform closest to their counterparts is ERPNext, which exposes only 7 tools. This suggests that a small number of general-purpose, flexible tools (ERPNext's API provides generic CRUD operations over all document types) can be more effective than a large catalog of narrow, specialized endpoints that may not cover the specific operations a task requires. Throughout this paper, \emph{tool-use agent} refers to the specific architecture we evaluate: an agent equipped with a large, curated set of narrow, per-endpoint tools exposed via an MCP server (Appendix~\ref{sec:mcp-servers}); this is a property of the tool catalogs we test, not an inherent limitation of the MCP protocol itself, which is agnostic to how many or how granular the exposed tools are. Tool-use agents are consistently the cheapest, confirming that their bottleneck is tool coverage rather than efficiency. We investigate this further in Appendix~\ref{sec:mcp-subset}, where we restrict the evaluation to the subset of tasks that are feasible for all three paradigms, and in Appendix~\ref{sec:api-call}, where we test this distinction directly with a single generic, MCP-style tool that closes most of the gap despite offering no more structure than one flexible API call. Even on the feasible-task subset alone, tool-use agents underperform due to rigid tool interfaces that limit which fields can be set and which query patterns can be expressed.

\textbf{Web agents offer more flexibility at a higher cost.} By interacting with the platform through a browser, web agents can in principle perform any action a human user would, and this flexibility translates to strong success rates: they achieve the highest or tied-highest accuracy in 8 of 12 platform--model combinations. However, this comes at a substantial cost premium. On ServiceNow, web agents cost 2--7$\times$ more than terminal agents across all four backbones, as each interaction processes large accessibility trees and screenshots. The overhead is especially pronounced for Opus 4.6 on ERPNext, where web agents cost \$6.49 per task compared to \$0.72 for the terminal agent: a 9$\times$ difference with only a modest accuracy gain (81.6\% vs.\ 76.8\%).

\textbf{Terminal agents offer the best cost--performance tradeoff.} Terminal agents match or exceed web agent accuracy in 8 of 12 platform--model combinations, treating scores within one standard error as ties, while consistently costing less, often by a factor of 5 and more. On the full benchmark the two paradigms fall within about a point of each other for three of the four models, so the terminal advantage over browser access is one of cost rather than accuracy; against tool-use agents it is a large accuracy gap on every platform as well. On ServiceNow, terminal-based coding agents achieve the highest success rate for all 4 models, although for Claude Sonnet 4.6 and Claude Opus 4.6 the margin over the web agent (1.2 and 1.5 points) is comparable to the run-to-run variation we report in Appendix~\ref{sec:reproducibility} and is not evidence of a paradigm-level difference. With Gemini 3.1 Pro the terminal agent reaches 77.5\% averaged across all environments at just \$0.09 per task, the most cost-efficient configuration in the table. Their advantage stems from programmatic interaction through shell commands and API calls, which avoids the overhead of UI rendering while retaining the flexibility to compose arbitrary operations that MCP tools may not expose. This ordering is not confined to frontier models: it also holds on every platform for two smaller open-weight backbones (Appendix~\ref{sec:open-weight}). We investigate where terminal agents still fail in Section~\ref{sec:cli-failures-discussion}.

\vspace{-7pt}
\subsection{Access to self-generated Skills}

% \begin{wrapfigure}{r}{0.5\textwidth}
\begin{figure}
\centering
% \vspace{-30pt}
\includegraphics[width=0.95\textwidth]{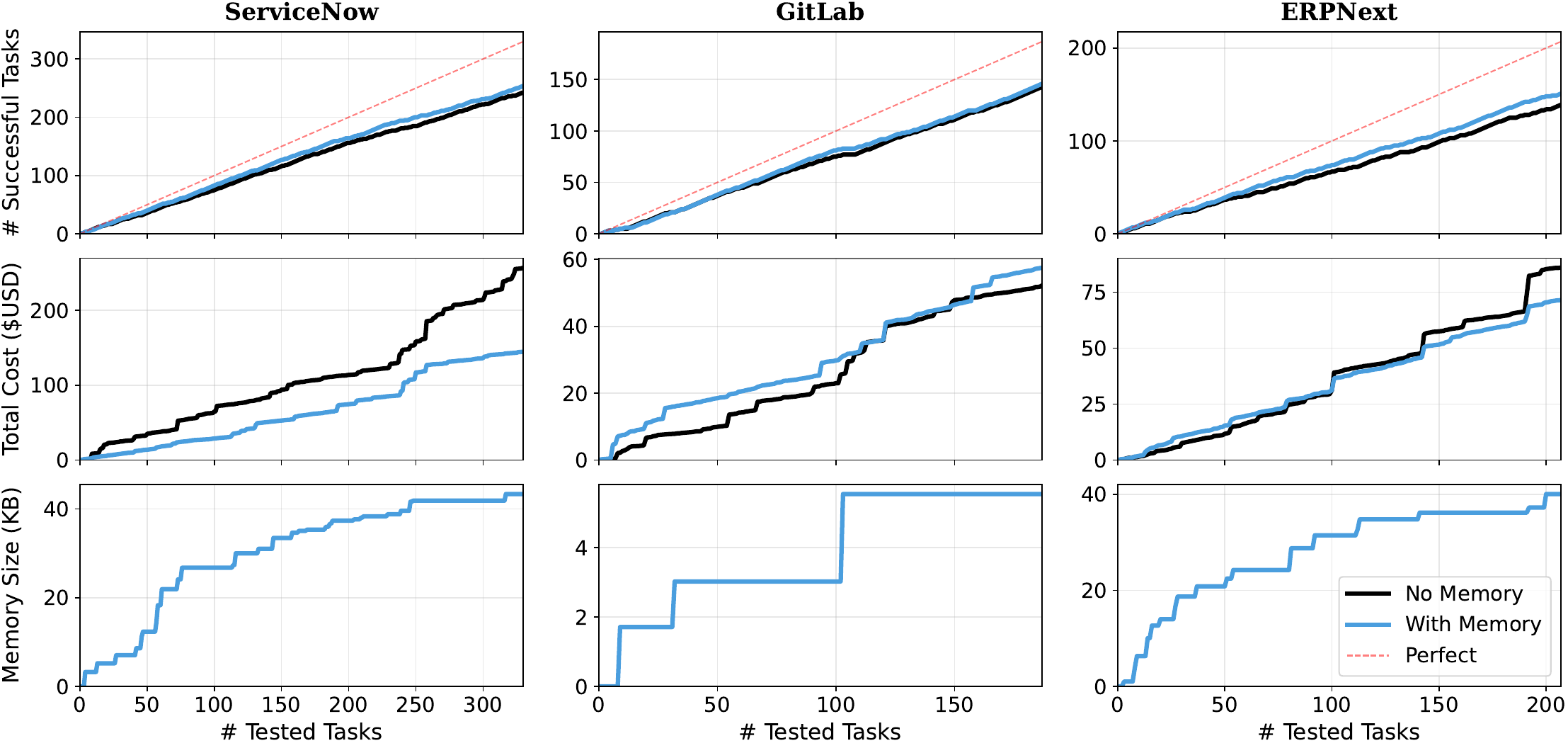}
% \vspace{-18pt}
\caption{\small{\textbf{Skills accumulation over sequential tasks.}
Top: cumulative number of successful tasks.
Middle: cumulative cost (\$USD).
Bottom: skills directory size (KB).
The agent with memory (blue) accumulates reusable procedures; the baseline (black) starts fresh every time.}}
\label{fig:skills-grid}
\vspace{-10pt}
% \end{wrapfigure}
\end{figure}

We investigate whether terminal agents can effectively ``learn on the job.'' As agents perform tasks over time, we allow them to store troubleshooting strategies, workflows, and other useful notes in a persistent ``skills'' directory, which they may organize freely. We sequentially go over the full dataset once and compare the cumulative number of tasks completed and cumulative cost of agents equipped with and without these skills over time. Results can be seen in Figure~\ref{fig:skills-grid}. Both configurations use Sonnet 4.6.

Persistent memory improves success rate across all three platforms, though the magnitude varies. The largest gain is on ERPNext (+5.3pp), where tasks frequently involve non-obvious field names and multi-step dependency chains that benefit from recorded procedures. ServiceNow improves by 3.4pp, while GitLab shows only a marginal gain (+1.6pp), consistent with its tasks being more straightforward API calls that the model can handle from parametric knowledge alone.

Beyond accuracy, the more striking effect is on cost. On ServiceNow, the agent with memory costs 43.7\% less per task on average (\$0.44 vs.\ \$0.78), and on ERPNext the reduction is 23.9\% (\$0.35 vs.\ \$0.46); on GitLab, by contrast, cost rises slightly (\$0.31 vs.\ \$0.28). One might expect memory accumulation to act as an investment, with a small upfront cost amortized over subsequent tasks, but in practice the overhead is modest: the agent checks the skills directory in 1--2 calls and writes new entries only on genuinely novel information, so maintenance is small relative to the savings from avoiding redundant API exploration.

Memory growth follows the same pattern. On ServiceNow and ERPNext, the skills directory grows rapidly over the first third of tasks as the agent records working procedures for new task types, then plateaus once most patterns have been observed. On GitLab it stays minimal: the agent writes to its skills directory only three times, producing just two skill files. The value of persistent memory therefore depends on how unfamiliar the platform's API is to the base model, with ERPNext, a less common platform, benefiting most and GitLab's well-documented REST API offering little the model does not already know. We discuss the types of skills agents generate in Section~\ref{sec:skills-discussion}.

Since each platform is traversed once in a fixed order with a single backbone, these magnitudes depend on the sequence encountered, as a task arriving early yields a skill later tasks can reuse. We therefore read them as evidence that run-time accumulation pays off, not as calibrated estimates of how much.

\subsection{API-First Agents with Browser Fallback}
\label{sec:hybrid}

We ask whether hybrid agents equipped with both a terminal and a Playwright browser can do \textit{more} than either tool alone.
Given that terminal agents strike the best balance between success rate and cost (Section~\ref{sec:agent_types}), but might in some cases be limited by the way they interact with the environment (Section~\ref{sec:cli-limitations}), we investigate whether adding browser access allows agents to handle tasks that are awkward to reach through API calls alone, without sacrificing the efficiency of programmatic interaction. Results are shown in Table~\ref{tab:results-hybrid}.

\begin{table}[ht]
\vspace{-5pt}
\centering
\small
\setlength{\tabcolsep}{6pt}
\begin{tabular}{l c cc c cc c cc c cc}
\toprule
&&\multicolumn{2}{c}{\textbf{ServiceNow}} &&
  \multicolumn{2}{c}{\textbf{GitLab}} &&
  \multicolumn{2}{c}{\textbf{ERPNext}} &&
  \multicolumn{2}{c}{\textbf{Overall}} \\
\textbf{Agent} && \multicolumn{2}{c}{(330)} &&
  \multicolumn{2}{c}{(192)} &&
  \multicolumn{2}{c}{(207)} &&
  \multicolumn{2}{c}{(729)} \\
\cmidrule(lr){3-4}
\cmidrule(lr){6-7}
\cmidrule(lr){9-10}
\cmidrule(lr){12-13}
&& \textit{SR~(\%)} & \textit{Cost~(\$)} &&
  \textit{SR~(\%)} & \textit{Cost~(\$)} &&
  \textit{SR~(\%)} & \textit{Cost~(\$)} &&
  \textit{SR~(\%)} & \textit{Cost~(\$)} \\
\midrule
\rowcolor{gray!10}\multicolumn{13}{c}{\textit{Claude Sonnet 4.6}} \\
Web            &&
    \textbf{72.4} & 4.49 &&
    \textbf{82.9} & 0.88 &&
    61.8 & 3.63 &&
    \textbf{72.2} & 3.29 \\
Terminal            &&
    \textbf{73.6} & \textbf{0.78} &&
    76.5 & \textbf{0.28} &&
    \textbf{67.6} & \textbf{0.46} &&
    \textbf{72.7} & \textbf{0.56} \\
Hybrid            &&
    \textbf{72.1} & 3.02 &&
    77.7 & 0.64 &&
    \textbf{66.7} & 1.44 &&
    \textbf{72.0} & 1.94 \\
\midrule
\rowcolor{gray!10}\multicolumn{13}{c}{\textit{Claude Opus 4.6}} \\
Web            &&
    77.6 & 4.21 &&
    \textbf{81.9} & 0.85 &&
    \textbf{81.6} & 6.49 &&
    79.9 & 3.97 \\
Terminal            &&
    79.1 & \textbf{1.94} &&
    \textbf{80.2} & \textbf{0.50} &&
    76.8 & \textbf{0.72} &&
    78.7 & \textbf{1.22} \\
Hybrid            &&
    \textbf{83.0} & 2.57 &&
    \textbf{80.3} & 0.94 &&
    \textbf{79.7} & 2.84 &&
    \textbf{81.4} & 2.22 \\
\bottomrule
\end{tabular}
\vspace{-5pt}

\caption{\textbf{Hybrid agent comparison.} Success rate and cost per task for web, terminal, and hybrid agents on three enterprise platforms. The hybrid agent has access to both a terminal and a Playwright-controlled browser. Bold indicates the best SR and lowest cost within each platform--model group, along with any score within one standard error of the best.}
\label{tab:results-hybrid}
\vspace{-5pt}
\end{table}

To quantify the opportunity, we measure the complementarity of terminal and web agents on ServiceNow: out of 330 tasks, Terminal solves 55 that Web cannot, and Web solves 51 that Terminal cannot. A perfect oracle selecting the better agent per task would achieve 89.1\%, more than 15 percentage points above either agent alone.
This gap is the theoretical ceiling for hybrid access.

With Sonnet 4.6, the hybrid agent does not close this gap. It achieves 72.1\% on ServiceNow, slightly below both Terminal (73.6\%) and Web (72.4\%), at a cost of \$3.02, four times higher than Terminal. With Opus 4.6, the picture improves: the hybrid agent reaches 83.0\% on ServiceNow, the highest of any configuration, while costing \$2.57, less than the web agent (\$4.21) but more than Terminal (\$1.94). Full results for all experiments reported above are given in Appendix~\ref{sec:full-results}, and additional experiments in Appendix~\ref{sec:additional-experiments}.

\section{Analysis and Discussion}

In this section, we perform error analysis to get a better qualitative understanding of how terminal agents operate. We build custom tooling to help us explore traces and compare runs (see Appendix~\ref{sec:tooling}).

\subsection{What problems do terminal agents still fail on?}\label{sec:cli-failures-discussion}

\subsubsection{Failure is not driven by tool unreliability}

To understand failure modes, we analyze the traces from our ServiceNow baseline
evaluation for Claude Sonnet 4.6 (Section~\ref{sec:agent_types}). In this setting, nearly all tool calls (97.3\%) are \texttt{curl} commands issued against the ServiceNow REST API, so tool call failures correspond almost entirely to API interaction errors. Table~\ref{tab:tool-call-outcome-taxonomy} in Appendix~\ref{sec:tool-call-outcome-taxonomy} presents an overview of the different tool call outcomes along with examples.

\begin{figure*}[h]
\centering
\begin{minipage}[t]{0.49\textwidth}
    \small
    \centering
    \includegraphics[width=\textwidth]{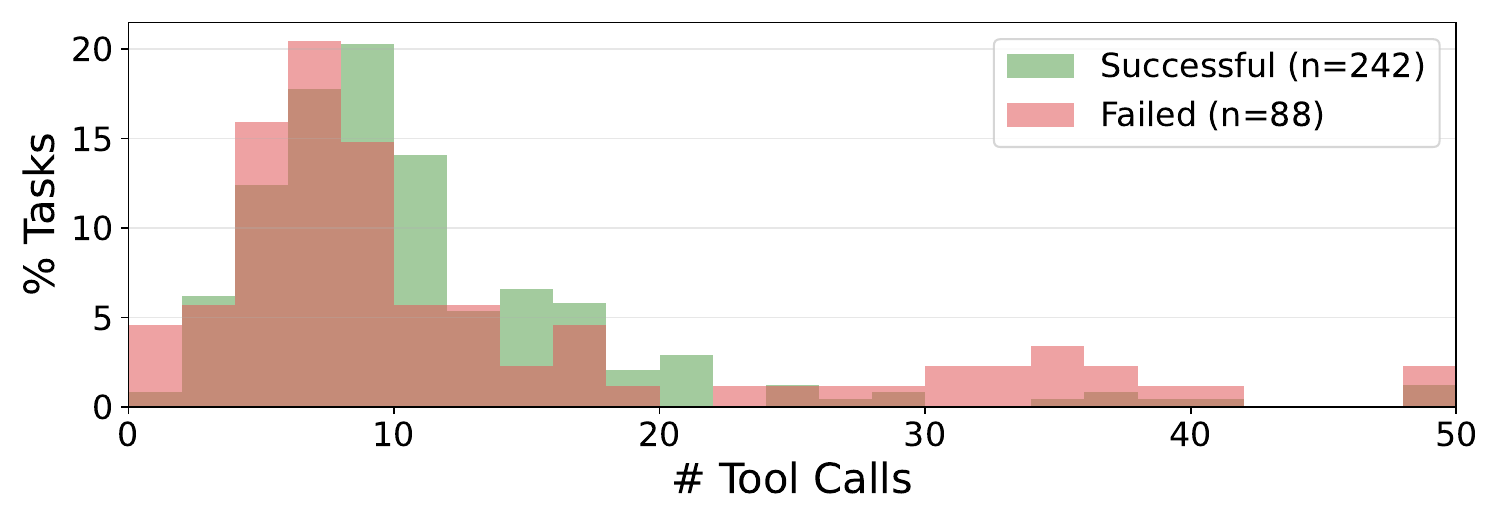}
    \vspace{-18pt}
    \captionof{figure}{\small{Tool calls per task for successful and failed tasks; those exceeding 50 are capped at 50.}}
    \label{fig:tool-calls-histogram}
\end{minipage}
\hfill
\begin{minipage}[t]{0.49\textwidth}
    \centering
    \small
    \includegraphics[width=\textwidth]{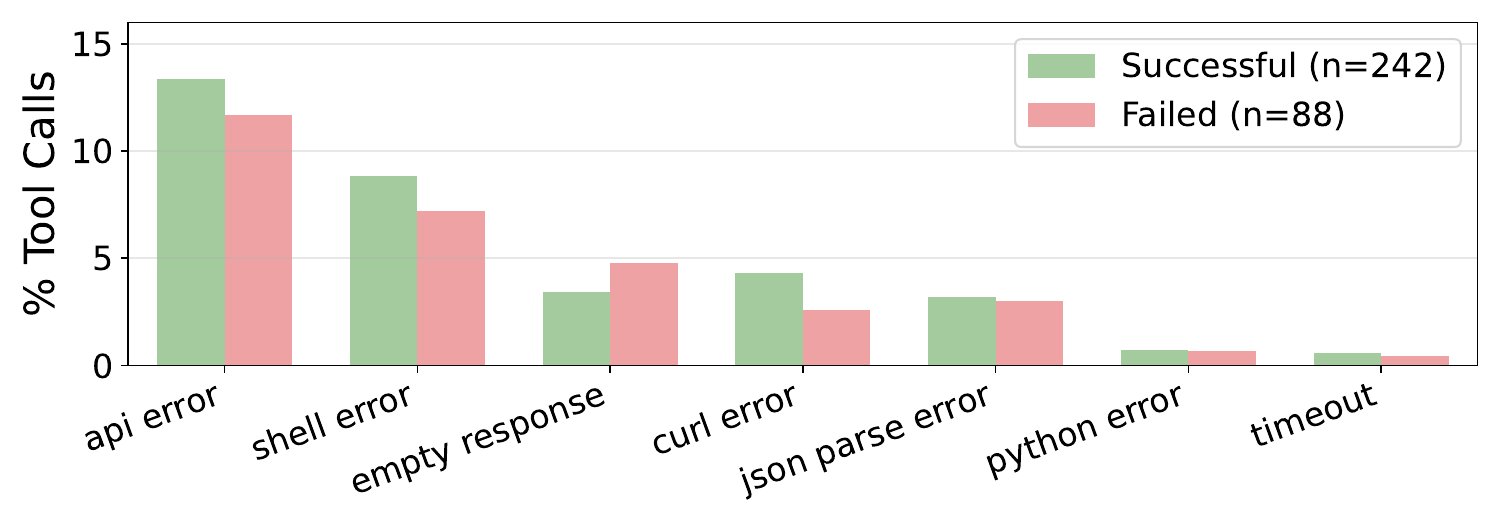}
    \vspace{-18pt}
    \captionof{figure}{\small{API error type breakdown as a percentage of total tool calls, comparing successful and failed tasks.}}
    \label{fig:api-error-breakdown}
\end{minipage}
% \vspace{-15pt}
\end{figure*}

Figure~\ref{fig:tool-calls-histogram} shows the distribution of tool calls per task. The histograms for successful and failed tasks are broadly similar, suggesting that task difficulty (not sheer volume of interaction) drives failure. However, two differences stand out: failed tasks show a heavier concentration near zero, indicating that some agents fail early without meaningful progress, and a slight bump above 30 tool calls, suggesting agents occasionally get stuck in unproductive loops before hitting the turn limit.

Figure~\ref{fig:api-error-breakdown} breaks down error types as a fraction of total tool calls in each group. Successful and failed tasks encounter comparable rates of API errors, \texttt{curl} failures, and shell errors, indicating that the agent generally recovers from individual tool call failures and that task-level failure comes from its inability to make progress rather than from unreliable tools.

\subsubsection{Where the browser is the shorter path}\label{sec:cli-limitations}

While terminal agents achieve the best cost--performance tradeoff, some tasks are markedly more convoluted to accomplish through API calls than through the UI. It is worth separating two cases: tasks a terminal agent can reach programmatically but only indirectly, and the narrower set for which the platform exposes no programmatic path at all.

Some operations are tied to the browser session, which makes them awkward to drive through API calls. ServiceNow's \emph{impersonation} feature, which lets an administrator act as another user, is a clear example. The terminal agent finds what appears to be the correct endpoint (\texttt{/api/now/ui/user/impersonate}) and receives an HTTP 200 response, but the impersonation never takes effect: the session state it depends on is carried by browser cookies rather than the response. Reproducing this programmatically requires capturing and replaying those cookies, which none of our agents discovered. The web agent completes the task in two clicks: the user menu, then the target user.

Tasks that target values rendered in the UI, such as extracting a figure from a dashboard chart, are reachable but indirect. The terminal agent queries the underlying tables and recomputes the aggregation, which usually works, but the rendered value may differ due to rounding, formatting, or chart-specific display logic. The web agent reads the rendered state, making a one-step observation of what is otherwise a multi-step reconstruction. That shorter path does not translate into a higher success rate: these are among the hardest tasks in our benchmark for either paradigm, and the web agent scores no better on them than the terminal agent. The browser's advantage here is directness instead of accuracy.

The one genuine capability gap arises with tasks involving complex UI interfaces such as drag-and-drop workflow editors. Consider a task outside our benchmark: \emph{``Create a workflow triggered daily that looks up all high priority incidents assigned to Fred Luddy and send him an email with the details.''} A web agent completes this through standard UI interactions in Flow Designer. While recent work shows such workflows can be generated programmatically~\citep{ayala2024generating}, this capability is not exposed through the platform's public APIs, leaving terminal agents no reliable path.

Appendix~\ref{sec:per-category} breaks these results down by category: the web agent wins the categories above, most others are ties, and the terminal agent spends fewer tokens almost everywhere. Browser access is therefore a shorter path for a minority of tasks rather than a capability the terminal lacks, and combining both interfaces lets agents pick the better mode per subtask, which we explore with a hybrid agent in Section~\ref{sec:hybrid}.

\subsection{Where does the terminal advantage come from?}
\label{sec:factor-decomposition}

Our ablations isolate the contributing factors. Tool coverage is the largest: restricting to tasks all three paradigms can express (Appendix~\ref{sec:mcp-subset}) removes part of the tool-use deficit, but rigid signatures still limit which fields and query patterns are reachable. Tool granularity accounts for most of the rest, as a single generic \texttt{api\_call} tool recovers nearly the entire gap (Appendix~\ref{sec:api-call}). Most of the advantage over curated tool use therefore comes from flexible API access rather than from the terminal itself. The terminal is one general way to provide it, and the same interface also covers documentation, skills, and multi-step scripting (Appendix~\ref{sec:api-call}). Error recovery and token overhead are not part of the explanation: error rates are comparable across successful and failed trajectories (Section~\ref{sec:cli-failures-discussion}), and tool-use agents are the cheapest paradigm of all.

Together these results follow the Bitter Lesson: general interfaces beat hand-engineered structure. Tool-use agents come closest to parity on ERPNext's seven generic operations and collapse on ServiceNow's 83 narrow ones, so more curation buys less. What the agent works out for itself carries no such penalty, and skills written at run time cut ServiceNow cost by 43.7\%. The advantage is also stable rather than model-specific, with the gap over tool-use agents ranging from 34.7 to 40.4 points across our four backbones.

\subsection{What skills do terminal agents create?}
\label{sec:skills-discussion}

We examine the skills directories accumulated during evaluation to understand what knowledge agents record and how this varies across platforms. Across all three platforms, the agent created 38 skill files: 11 for ServiceNow, 2 for GitLab, and 25 for ERPNext. Every skill was marked as \texttt{verified}, indicating the agent confirmed each procedure on at least one subsequent task.

Skills are overwhelmingly procedural, presenting step-by-step recipes for creating or modifying records via the platform API. However, the bulk of each skill's content is not the procedure itself, which is typically a single API call, but rather the \emph{pitfalls} and \emph{field mappings} embedded within it. These capture non-obvious knowledge discovered through trial and error: mappings between UI labels and API fields, valid field values, which endpoints exist, and workarounds for issues such as shell quoting.

The nature of skills varies markedly across platforms. On \textbf{ServiceNow}, skills are larger and more encyclopedic (3.9\,KB on average), accumulating field mappings, identifiers, query patterns, and platform-specific quirks across tasks.

On \textbf{ERPNext}, skills are more numerous but smaller (25 files averaging 1.8\,KB), each focused on a single entity type. While the API structure is uniform, skills record subtle schema inconsistencies and side effects that can lead to silent failures, such as fields like \texttt{so\_required=1} having inverted semantics (allowing invoices \emph{without} sales orders).

On \textbf{GitLab}, only two skill files are created, both capturing non-obvious API conventions: the undocumented template name \texttt{plainhtml} for project creation and the access level mapping for member invitations. This scarcity suggests stronger parametric knowledge of the platform, as most tasks succeed without new skills.

A recurring pattern across platforms is the documentation of \textbf{shell quoting issues} with \texttt{eval curl} and JSON payloads: skills consistently warn against inline JSON and recommend temporary files or scripts instead, an infrastructure-level lesson that generalizes beyond individual tasks. Appendix~\ref{sec:skills-case-study} shows an agent leveraging skills to recall a previously discovered solution.

\subsection{How do hybrid agents choose their tools?}
\label{sec:hybrid-discussion}

To understand these results, we examine the hybrid agent's tool selection behavior across all 330 ServiceNow tasks with Sonnet 4.6. Overall, 82\% of tool calls are browser-based, indicating that the agent strongly favors the browser despite prompt guidance to prefer the terminal for data operations. However, this aggregate masks a clear per-category pattern: the agent uses the terminal almost exclusively for record creation (incidents, change requests, users) and the browser for everything else.

The browser dominance is not always a problem: for tasks more directly reached through the UI the hybrid agent correctly relies on it. On impersonation tasks the terminal agent scores 0\%, for the reason above, while the hybrid agent succeeds by navigating the user menu. For dashboard chart reading it combines both, locating the dashboard via the \texttt{pa\_dashboards} table, then switching to the browser with the resolved URL to read the chart.

Inefficiencies arise when the agent defaults to the browser for tasks that the terminal agent handles more efficiently. On filtering tasks, the hybrid agent spends 30 to 50 browser calls manipulating the filter UI, frequently hitting the turn limit, while the terminal agent constructs the equivalent \texttt{sysparm\_query} URL in a few API calls. Sorting and catalog ordering follow the same pattern: the browser path works, but far more expensively.

These results suggest that hybrid access is most valuable when two conditions hold: (1) the base model is capable enough to select the right tool for each subtask, and (2) the platform contains a meaningful proportion of tasks for which the browser is substantially more direct. \textit{Skills} that encode routing guidance may help the agent better balance terminal and browser use.

Together, these results support a simple practical principle for interface selection: default to a terminal-based, API-first agent, and escalate to the browser only for the classes of tasks identified in Section~\ref{sec:cli-limitations}, which captures most of the practical value of hybrid access (Section~\ref{sec:hybrid}) without paying the browser's cost on tasks the terminal agent already handles.

For platform providers, it suggests a matching rule: anything a user can do through the interface should also be doable through the API. Coding agents readily discover and compose whatever a platform exposes programmatically, so parity between the two surfaces is the most direct way to make a platform automatable.

\section{Conclusion}

This work examined whether sophisticated agent stacks are necessary for practical enterprise automation. Across benchmarks spanning multiple production-grade platforms, minimal coding agents that operate through a terminal and filesystem and interact directly with platform APIs match or outperform more complex architectures, including GUI-driven web agents and MCP-based tool-augmented agents, while remaining competitive on efficiency. This simplicity is deliberate: code and API calls keep the task close to the modality LLMs are most heavily trained on. The tasks we use to substantiate this are simple, reachable in a handful of API calls on one platform, but the trend across platforms, backbones, and task categories is consistent. More broadly, enterprise automation may benefit more from exposing stable programmable interfaces than from introducing additional abstraction layers, and a terminal-based, API-first agent should be the default baseline whenever a platform exposes a sufficiently expressive API. Browser access remains useful where the UI offers a more direct route, but as an extension to this baseline rather than the interface agents are built around.

Several directions remain for future work. Enterprise automation frequently requires long-horizon agents that coordinate across platforms, maintain state over extended interactions, and incorporate human oversight. Benchmarks with these properties, and coverage of verticals such as IT operations, HR, security, and finance, will be essential for understanding how agent architectures scale.

\section*{Limitations}

\paragraph{Scope of the central claim.} Our claim is that a terminal agent should be the default foundation for enterprise automation, not that every task is equally easy to solve with a terminal or a browser: some effects are one click away in a browser and ten API calls away in a terminal, because they live in the browser session, in information the interface renders rather than stores, or behind a UI affordance no endpoint mirrors. Our benchmark contains all three cases: 10 of the 330 ServiceNow tasks require a session-based impersonation flow, 40 target values rendered in dashboard charts that a terminal agent must reconstruct from the underlying tables, and GitLab's social actions are single clicks against endpoints whose identifiers must first be resolved. Terminal agents still lead on ServiceNow for all four backbones with them included. Separating them out (Appendix~\ref{sec:per-category}) leaves a narrower result than the aggregates suggest: the two paradigms are at parity on most task categories, while the terminal agent's accuracy advantage sits in operations that must be composed rather than clicked and its cost advantage holds nearly everywhere. The tasks themselves also condition the claim, being simple by the standards of what agents are asked to do today: most are reachable in a handful of API calls, require little exploration, and touch a single system. The trend across platforms, backbones, and task categories is nonetheless consistent, and nothing in the task distribution favors programmatic access, since our ServiceNow and GitLab tasks are adapted from WorkArena and WebArena, benchmarks built to evaluate web agents. We therefore expect the conclusion to carry to more complex workloads, and treat demonstrating it there as future work. The one genuine capability gap we identify, artifacts authored through complex UI tooling such as Flow Designer workflows (Section~\ref{sec:cli-limitations}), is uncommon in practice and does not appear in our task set.

\paragraph{Benchmark scope.} Our evaluation spans three production-grade platforms and 729 single-session tasks, but does not cover long-horizon workflows unfolding over many sessions, coordination across platforms within one task, or scenarios requiring human oversight. Such benchmarks are difficult to build: they need fully functioning applications where GUI, API, and tool-based agents can all be evaluated on the same tasks, which benchmarks exposing only database access or simplified tool interfaces cannot support.

\paragraph{Tool-use comparison scope.} Our tool-use results characterize the off-the-shelf MCP servers we evaluated, which expose 7 to 83 tools (Appendix~\ref{sec:mcp-servers}), rather than the full space of MCP implementations. Appendices~\ref{sec:mcp-subset} and~\ref{sec:api-call} isolate the effects of task coverage and tool granularity, and a field-complete server could narrow the gap further.

\paragraph{Outcome quality beyond success rate.} Our validators check that the requested state exists, not that nothing else changed. Appendix~\ref{sec:side-effects} audits every state-changing request. We find that unintended writes are rare for both paradigms. Measuring failure severity properly needs validators that assert the absence of unrequested changes.

\paragraph{Hybrid agent routing.} Our hybrid agent (Section~\ref{sec:hybrid}) routes between terminal and browser on the base model's implicit judgment, over-relying on the browser relative to the oracle upper bound reported there. We do not evaluate learned or rule-based routing policies.

\bibliography{main}
\bibliographystyle{tmlr}

\appendix

\clearpage

\section{Reproducibility Statement}\label{sec:reproducibility}

The full evaluation framework, evaluation datasets, environments, prompts, and documentation are publicly available through the StarShell GitHub repository (\url{https://github.com/ServiceNow/StarShell}) and Hugging Face collection (\url{https://huggingface.co/collections/ServiceNow-AI/starshell}). All prompts used in our experiments are included in Appendix~\ref{sec:prompts} and in the source code. Task definitions, success criteria, and evaluation procedures are fully specified in the evaluation framework, and all evaluation metrics and scoring scripts are deterministic given the model outputs.

\paragraph{Uncertainty and repeated runs.} Exhaustively repeating the full evaluation matrix was not feasible given the number of models, platforms, agent paradigms, and tasks, together with frontier-model API costs, rate limits, and the long wall-clock time required to execute complete agent trajectories. One complete pass over three paradigms, four backbones and three platforms costs approximately \$8{,}650 in inference, of which the web agent accounts for \$6{,}115, so ten repetitions of it would cost roughly \$86{,}500 and 87{,}480 agent trajectories. We therefore report two complementary measures of uncertainty.

First, because task outcomes are binary, we estimate uncertainty due to the finite benchmark size using the standard error of a sample proportion, $\sqrt{\hat{p}(1-\hat{p})/n}$, where $\hat{p}$ is the observed success rate and $n$ is the number of evaluated tasks. For success rates near 80\%, this corresponds to standard errors of approximately 2.2, 2.9, and 2.8 percentage points on ServiceNow, GitLab, and ERPNext, respectively, and approximately 1.5 percentage points over the full 729-task benchmark.

Second, we conducted targeted repeated-run evaluations to measure variation arising from stochastic model generation, tool execution, and environment interaction. We repeated the full ServiceNow evaluation with Gemini~3.1~Pro and the full GitLab evaluation with GPT-5.4 Thinking (Medium) across 3 independent runs for each agent paradigm. On ServiceNow, the standard deviation of success rate ranged from 0.5 to 1.4 percentage points across tool-use, web, and terminal agents; on GitLab, it ranged from 1.9 to 2.2 percentage points. The corresponding standard deviations in average inference cost ranged from \$0.005 to \$0.043 per task on ServiceNow and from \$0.019 to \$0.062 on GitLab. These results indicate that the principal terminal-versus-web and terminal-versus-tool-use comparisons are stable across repeated executions on the tested platform--model combinations. Nevertheless, the repeated-run analysis covers only a subset of the complete evaluation matrix, and small differences between methods should still be interpreted cautiously.

Both measures concern the mean success rate. A separate question is whether an agent handles \emph{every} instance of a recurring task. Appendix~\ref{sec:consistency} addresses this using the 10 seeds we generate per ServiceNow task type. Each seed varies the records and field values of the same operation, so the seeds give one proxy for pass\textasciicircum{}$k$ at no additional inference cost.

Unless stated otherwise, all models are evaluated using their default sampling parameters. For the repeated-run experiments described above, each run starts from the same initial task state but uses an independent model execution. Experiments are executed through the LiteLLM framework, using Azure endpoints for GPT-5 models, Vertex AI for Gemini models, and Amazon Bedrock endpoints for Claude models. All agent runs are executed in fully containerized environments to prevent cross-task contamination; after each task execution, the environment is reset to its original state to ensure a clean and identical starting condition. Outside the skills experiment, where the skills directory is carried forward by design, no state persists across tasks. Since decoding is stochastic, re-running an agent reproduces a distribution of trajectories rather than any specific one. What remains reproducible at the trajectory level is the analysis: per-step traces and agent and environment logs are recorded for every run, and the tooling used to inspect them (Appendix~\ref{sec:tooling}) is included in the release.

\section{Ethics and Safety Considerations}\label{sec:ethics}

Terminal-based agents operate with broad execution capabilities, including access to a filesystem and the ability to issue live API calls to enterprise platforms. In our setting, the filesystem is scoped to the agent and does not directly expose user data, limiting risks from local state. However, the agent can directly read and modify records through platform APIs, which introduces the possibility of unintended or harmful actions. In practice, the action space the agent exercises is far narrower than the one it is granted: in our ServiceNow error analysis (Section~\ref{sec:cli-failures-discussion}), 97.3\% of tool calls are \texttt{curl} invocations against the REST API, with the agent using the shell largely as an authenticated HTTP client. This is why we regard the credential, rather than the command line, as the meaningful control point. Allowlisting shell commands is a weak constraint, since a single permitted \texttt{curl} re-exposes the entire API surface and forbidding it only redirects the agent to a short script, whereas a role-scoped service account, read-only credentials for retrieval-only workloads, and platform-side auditing bound what any command can accomplish.

Importantly, these risks are not unique to terminal-based agents, but arise more generally in autonomous systems that interact with real-world services. Compared to curated tool interfaces or GUI-based agents, terminal agents expose a more expressive but less constrained action space, and we do not dispute that typed tool interfaces retain genuine advantages in schema constraints, permission control, and auditability that does not depend on interpreting shell commands. Those advantages hold, however, only while the catalog covers the work: on 12 ServiceNow tasks our tool-use agent reached data no tool exposed by creating a script include and running server-side code (Appendix~\ref{sec:side-effects}), so the narrower interface redirected the action into a less auditable form rather than preventing it. We would argue, though, that most of this gap can be closed at the environment boundary rather than by narrowing the agent's action vocabulary: an unprivileged container with a workspace-scoped filesystem, network egress restricted to the platform endpoint, and a local proxy that holds the credential and attaches it to outbound requests, which keeps the secret out of the agent's context and provides a single point for request-level allowlisting and logging. Our own harness does not yet do this, injecting authentication headers as environment variables the agent can read (Appendix~\ref{sec:prompts}). Persisted skills deserve separate care, since they are written by the agent and re-read by later tasks, and are better treated as untrusted input scoped to a single trust boundary than as trusted memory.

In this work, we focus on identifying a minimal and effective interaction abstraction for enterprise automation, and we view safety, reliability, and access control as complementary layers built on top of it. We have argued for a particular placement of those layers rather than evaluated them: we run no adversarial tasks, and the audit of unintended state changes in Appendix~\ref{sec:side-effects} measures the requests agents issue rather than the state the platform is left in. Our results should accordingly be read as evidence about capability and cost instead of as establishing that terminal agents are ready for unsupervised operation on production systems.

\section{Additional Experiments}\label{sec:additional-experiments}

\subsection{Fairness of paradigm comparison}\label{sec:mcp-subset}

The results in Table~\ref{tab:results-main} include tasks that are structurally infeasible for tool-use agents due to missing tools, which might inflate the gap between paradigms.
For instance, the ServiceNow MCP server exposes no tool for ordering from the service catalog, navigating to a specific page, reading a dashboard chart, impersonating a user, or operating on hardware assets or problem records, all of which are checkable against Table~\ref{tab:servicenow-mcp-tools}. Together these account for 210 of the 330 ServiceNow tasks, just under two thirds of that benchmark.
Similarly, the ERPNext MCP server exposes create, read and update operations but no delete, which excludes exactly its ten deletion tasks (Table~\ref{tab:erpnext-mcp-tools}); and the GitLab MCP server has no tool for following a user, starring a project, or adding a project member, and in its default configuration does not expose its milestone tool group, so the social and milestone task types are unreachable (Table~\ref{tab:gitlab-mcp-tools}).

To provide a fairer comparison, Table~\ref{tab:results-subset} restricts evaluation to the subset of task types that are feasible for all three paradigms.
We retain 12 of 33 ServiceNow task types (120 instances), 15 of 28 GitLab task types (127 instances), and 78 of 82 ERPNext task types (197 instances).

\begin{table}[ht]
\centering
\small
\setlength{\tabcolsep}{6pt}
\begin{tabular}{l c cc c cc c cc c cc}
\toprule
&&\multicolumn{2}{c}{\textbf{ServiceNow}} &&
  \multicolumn{2}{c}{\textbf{GitLab}} &&
  \multicolumn{2}{c}{\textbf{ERPNext}} &&
  \multicolumn{2}{c}{\textbf{Overall}} \\
\textbf{Agent} && \multicolumn{2}{c}{(120)} &&
  \multicolumn{2}{c}{(127)} &&
  \multicolumn{2}{c}{(197)} &&
  \multicolumn{2}{c}{(444)} \\
\cmidrule(lr){3-4}
\cmidrule(lr){6-7}
\cmidrule(lr){9-10}
\cmidrule(lr){12-13}
&& \textit{SR~(\%)} & \textit{Cost~(\$)} &&
  \textit{SR~(\%)} & \textit{Cost~(\$)} &&
  \textit{SR~(\%)} & \textit{Cost~(\$)} &&
  \textit{SR~(\%)} & \textit{Cost~(\$)} \\
\midrule
\rowcolor{gray!10}\multicolumn{13}{c}{\textit{Claude Sonnet 4.6}} \\
Tool-use            &&
    24.2 & 1.30 &&
    66.9 & 0.52 &&
    58.4 & \textbf{0.14} &&
    51.6 & 0.56 \\
Web            &&
    66.7 & 4.57 &&
    \textbf{86.4} & 0.91 &&
    60.7 & 3.82 &&
    69.6 & 3.20 \\
Terminal            &&
    \textbf{75.8} & \textbf{0.49} &&
    80.2 & \textbf{0.44} &&
    \textbf{66.0} & 0.48 &&
    \textbf{72.7} & \textbf{0.47} \\
\midrule
\rowcolor{gray!10}\multicolumn{13}{c}{\textit{Claude Opus 4.6}} \\
Tool-use            &&
    30.0 & \textbf{0.62} &&
    69.3 & 1.10 &&
    72.4 & \textbf{0.18} &&
    60.0 & \textbf{0.56} \\
Web            &&
    70.0 & 4.15 &&
    \textbf{82.7} & 0.87 &&
    \textbf{81.5} & 6.86 &&
    \textbf{78.7} & 4.40 \\
Terminal            &&
    \textbf{80.8} & 1.78 &&
    \textbf{81.9} & \textbf{0.63} &&
    75.6 & 0.75 &&
    \textbf{78.8} & 0.99 \\
\midrule
% \rowcolor{gray!10}\multicolumn{13}{c}{\textit{GPT-5.2 (Medium)}} \\
% Tool-use            &&
%     36.7 & \textbf{0.08} &&
%     70.1 & 0.26 &&
%     72.1 & \textbf{0.20} &&
%     61.9 & \textbf{0.18} \\
% Web            &&
%     73.1 & 0.60 &&
%     \textbf{76.2} & 0.20 &&
%     \textbf{75.6} & 1.39 &&
%     \textbf{75.1} & 0.84 \\
% Terminal            &&
%     \textbf{83.3} & 0.14 &&
%     74.0 & \textbf{0.17} &&
%     69.5 & 0.50 &&
%     \textbf{74.5} & 0.31 \\
% \midrule
\rowcolor{gray!10}\multicolumn{13}{c}{\textit{GPT-5.4 Thinking (Medium)}} \\
Tool-use            &&
    39.2 & \textbf{0.11} &&
    70.1 & 0.35 &&
    66.0 & \textbf{0.19} &&
    59.9 & 0.21 \\
Web            &&
    75.0 & 0.19 &&
    \textbf{85.0} & 0.17 &&
    \textbf{71.6} & 0.39 &&
    \textbf{76.4} & 0.27 \\
Terminal            &&
    \textbf{86.7} & \textbf{0.11} &&
    74.0 & \textbf{0.11} &&
    \textbf{69.0} & 0.24 &&
    \textbf{75.2} & \textbf{0.17} \\
\midrule
\rowcolor{gray!10}\multicolumn{13}{c}{\textit{Gemini 3.1 Pro}} \\
Tool-use            &&
    26.7 & 0.08 &&
    71.2 & 0.17 &&
    66.0 & \textbf{0.08} &&
    56.8 & 0.11 \\
Web            &&
    54.2 & 0.32 &&
    \textbf{89.0} & 0.24 &&
    \textbf{73.3} & 1.35 &&
    72.6 & 0.75 \\
Terminal            &&
    \textbf{81.7} & \textbf{0.05} &&
    83.3 & \textbf{0.06} &&
    \textbf{73.0} & 0.11 &&
    \textbf{78.3} & \textbf{0.08} \\
\bottomrule
\end{tabular}
\caption{\textbf{Results on the feasible task subset for tool-use agents.} Success rate (SR, $\uparrow$) and average cost per task ($\downarrow$) restricted to task types that are feasible for all three agent paradigms.
Bold indicates the best SR and lowest cost within each platform--model group, along with any score within one standard error of the best.
ServiceNow: 12 task types (120 instances), GitLab: 15 task types (127 instances), ERPNext: 78 task types (197 instances).}
\label{tab:results-subset}
\end{table}

On this subset, tool-use agents improve substantially: Sonnet tool-use on ServiceNow rises from 11.5\% to 24.2\%, and the overall gap narrows across all models.
However, terminal and web agents still outperform tool-use agents consistently. Terminal agents achieve the highest overall success rate in 3 of 4 model configurations (tied with web agents overall with Claude Opus 4.6), while remaining the most cost-efficient in 3 out of 4 cases.
The remaining tool-use gap on feasible tasks can be attributed to the fact that even when a tool exists for a given operation, its interface often restricts which fields can be set or updated. For example, a \texttt{create\_incident} tool may expose only a handful of parameters while the underlying API accepts dozens of fields. When a task requires setting a field that the tool does not expose, the tool-use agent has no recourse, whereas terminal and web agents can construct arbitrary payloads or interact with any form element.

This points to a fundamental limitation of curated, multi-tool catalogs specifically, not of MCP as an interaction paradigm: \textbf{an agent restricted to a fixed catalog of narrow, per-endpoint tools can only perform operations that catalog explicitly exposes, and only with the parameters each tool defines}. Investing in better MCP servers (e.g. adding more tools, exposing more fields) can narrow the gap, but the constraint remains structural: every narrow tool is a rigid interface over an API that the terminal agent can access directly and flexibly. A terminal agent facing an unfamiliar endpoint can explore, adapt its payloads, and even write scripts to automate complex sequences, capabilities that no fixed tool registry can fully replicate.
MCP servers built as large catalogs of narrow tools are ultimately bounded by the same API surface they wrap, but present it through many rigid, per-operation interfaces rather than one flexible one. Appendix~\ref{sec:api-call} shows that replacing this catalog with a single generic tool recovers most of the gap, confirming that tool granularity, rather than the MCP mechanism itself, drives this result.

\subsection{Disentangling Tool Granularity from Terminal Access}\label{sec:api-call}

The comparison above conflates two factors that could each explain the terminal agent's advantage over tool-use agents: (1) access to a single, flexible interface for issuing arbitrary API requests, rather than a curated registry of narrow, per-endpoint tools, and (2) access to a filesystem and shell, which additionally enables browsing documentation, saving and reusing Skills, and scripting multi-step operations. To isolate the first factor, we introduce an \textbf{API-call agent}: an agent equipped with a single structured tool, \texttt{api\_call(method, path, body, query)}, that issues an authenticated HTTP request to the platform's REST API and returns the response. Authentication is handled automatically, exactly as with our MCP servers, but the agent can invoke any method, path, query, and body rather than being restricted to a fixed catalog of predefined operations. Unlike the terminal agent, the API-call agent has no filesystem, no shell, and no ability to browse documentation, save Skills, or chain commands: from the model's perspective it is a single, maximally flexible tool call, structurally analogous to an MCP tool but without a curated schema per endpoint.

Although its parameter schema is generic in shape, each deployment of the API-call tool is still bound to a single platform's base URL and injected credentials at configuration time, so supporting a new platform requires instantiating a new tool binding, conceptually one tool \emph{per platform} rather than one tool for all platforms. This differs from the terminal agent, whose single shell tool requires no new tool code to support an additional platform: cross-platform generality there comes entirely from configuration, environment variables, Skills, and documentation, rather than from re-implementing the tool itself.

We evaluate the API-call agent with Claude Sonnet 4.6 and Claude Opus 4.6 across all three platforms and the full 729-task benchmark. Table~\ref{tab:results-api-call} reports these results alongside the tool-use, web, and terminal agents for the same two models, reproduced from Table~\ref{tab:results-main}.

\begin{table}[ht]
\centering
\small
\setlength{\tabcolsep}{6pt}
\begin{tabular}{l c cc c cc c cc c cc}
\toprule
&&\multicolumn{2}{c}{\textbf{ServiceNow}} &&
  \multicolumn{2}{c}{\textbf{GitLab}} &&
  \multicolumn{2}{c}{\textbf{ERPNext}} &&
  \multicolumn{2}{c}{\textbf{Overall}} \\
\textbf{Agent} && \multicolumn{2}{c}{(330)} &&
  \multicolumn{2}{c}{(192)} &&
  \multicolumn{2}{c}{(207)} &&
  \multicolumn{2}{c}{(729)} \\
\cmidrule(lr){3-4}
\cmidrule(lr){6-7}
\cmidrule(lr){9-10}
\cmidrule(lr){12-13}
&& \textit{SR~(\%)} & \textit{Cost~(\$)} &&
  \textit{SR~(\%)} & \textit{Cost~(\$)} &&
  \textit{SR~(\%)} & \textit{Cost~(\$)} &&
  \textit{SR~(\%)} & \textit{Cost~(\$)} \\
\midrule
\rowcolor{gray!10}\multicolumn{13}{c}{\textit{Claude Sonnet 4.6}} \\
Tool-use            &&
    11.5 & 0.76 &&
    45.2 & 0.48 &&
    55.6 & 0.14 &&
    32.9 & 0.51 \\
Web            &&
    \textbf{72.4} & 4.49 &&
    \textbf{82.9} & 0.88 &&
    61.8 & 3.63 &&
    \textbf{72.2} & 3.29 \\
Terminal            &&
    \textbf{73.6} & 0.78 &&
    76.5 & 0.28 &&
    \textbf{67.6} & 0.46 &&
    \textbf{72.7} & 0.56 \\
API            &&
    \textbf{72.6} & \textbf{0.17} &&
    \textbf{80.8} & \textbf{0.17} &&
    \textbf{68.1} & \textbf{0.11} &&
    \textbf{73.5} & \textbf{0.15} \\
\midrule
\rowcolor{gray!10}\multicolumn{13}{c}{\textit{Claude Opus 4.6}} \\
Tool-use            &&
    16.1 & 0.66 &&
    46.8 & 0.90 &&
    68.9 & \textbf{0.17} &&
    39.2 & 0.58 \\
Web            &&
    \textbf{77.6} & 4.21 &&
    \textbf{81.9} & 0.85 &&
    \textbf{81.6} & 6.49 &&
    \textbf{79.9} & 3.97 \\
Terminal            &&
    \textbf{79.1} & 1.94 &&
    \textbf{80.2} & 0.50 &&
    76.8 & 0.72 &&
    \textbf{78.7} & 1.22 \\
API            &&
    \textbf{79.5} & \textbf{0.22} &&
    \textbf{80.3} & \textbf{0.31} &&
    77.8 & 0.25 &&
    \textbf{79.2} & \textbf{0.25} \\
\bottomrule
\end{tabular}
\caption{\textbf{Isolating tool granularity from terminal access.} Success rate (SR, $\uparrow$) and average cost per task ($\downarrow$) for tool-use, web, terminal, and API-call agents. The API-call agent uses a single generic, authenticated HTTP tool with no filesystem or shell access. Tool-use, web, and terminal rows are reproduced from Table~\ref{tab:results-main}. Bold indicates the best SR and lowest cost within each platform--model group, along with any score within one standard error of the best.}
\label{tab:results-api-call}
\end{table}

The API-call agent closes nearly all of the gap between tool-use and terminal agents, and in several cases matches or exceeds the terminal agent outright. With Sonnet~4.6, it achieves the highest overall success rate of any paradigm (73.5\%, vs.\ 72.7\% for terminal and 32.9\% for tool-use) while remaining the cheapest (\$0.15, vs.\ \$0.56 for terminal and \$0.51 for tool-use). With Opus~4.6, it reaches 79.2\% overall, within 0.7 points of web (79.9\%) and above terminal (78.7\%), again at a fraction of the cost of either (\$0.25, vs.\ \$1.22 for terminal and \$3.97 for web), and even cheaper than tool-use (\$0.58).

We emphasize that the API-call agent is a controlled ablation designed to isolate the tool-granularity variable, not a proposed replacement for the terminal agent. These results indicate that a substantial part of the terminal agent's advantage over tool-use is driven by the flexibility of unconstrained API access rather than by terminal or filesystem access per se: a single tool, deployed one per platform, that removes the curated-schema constraint recovers most of the gap. However, this comparison also removes several capabilities that remain practically important and that our benchmark's task profile likely understates.

First, writing to a filesystem lets the terminal agent offload large intermediate results, such as paginated table dumps or bulk query responses, outside the model's context window, and filter or transform them locally with standard utilities (\texttt{grep}, \texttt{jq}, \texttt{awk}) before feeding a distilled result back to the model. The API-call agent has no such scratch space: every response must be consumed directly in-context, which is unlikely to scale as gracefully to tasks involving larger data volumes than those in our benchmark.

Second, each \texttt{api\_call} invocation corresponds to one model-driven tool call, whereas a shell script can express a bulk operation over many records, such as updating a field across dozens of records, as a single loop within one tool call. Terminal agents in our benchmark already average 8.8--14.5 tool calls per task across platforms for Sonnet~4.6 and Opus~4.6 (Table~\ref{tab:results-full}); the API-call agent would need one additional model turn per record for the equivalent bulk operation. Our benchmark's tasks do not stress this difference heavily, so we would expect it to scale less favorably in both cost and latency on tasks requiring many similar operations in sequence.

Third, the API-call tool's fixed schema covers conventional JSON REST requests, but not other protocols a platform operation might require, such as multipart file uploads or streaming responses, which a shell can handle by invoking the appropriate CLI tool or client library directly.

A fourth consideration is scale of the API surface itself. Large enterprise platforms expose hundreds or thousands of endpoints, and exposing each as a separate tool creates either substantial context bloat or the need for a retrieval mechanism to select among tools before the task can begin, whereas an agent that discovers endpoints on demand incurs no such cost and can search, inspect, and iterate over the surface directly. Our platforms expose 7 to 83 tools, so we do not measure this regime and offer the point as a structural argument rather than an empirical result.

We therefore read these results as evidence that tool granularity, rather than terminal access, explains most of the gap between tool-use and terminal agents in Table~\ref{tab:results-main}, while the terminal agent's filesystem-backed intermediate storage, batched execution, and broader protocol support remain independently useful capabilities that this ablation does not exercise. The residual advantage of the terminal and hybrid agents on specific platforms and task types (Sections~\ref{sec:cli-limitations} and~\ref{sec:hybrid}) reflects exactly these capabilities, such as documentation retrieval, persistent Skills, and multi-step scripting.

\subsection{Generalization to open-weight models}
\label{sec:open-weight}

Our main experiments use frontier proprietary models, raising the possibility that the advantage of terminal agents depends on the strong coding and API-interaction capabilities of these models.
To test whether the conclusions generalize beyond frontier models, we evaluate two smaller open-weight models, Gemma-4-31B-it~\citep{gemmateam2026gemma4technicalreport} and Qwen-3.6-27B~\citep{yang2025qwen3technicalreport}, across the same three interaction paradigms and full set of 729 benchmark tasks.

\begin{table}[ht]
\centering
\small
\setlength{\tabcolsep}{6pt}
\begin{tabular}{l c cc c cc c cc c cc}
\toprule
&&\multicolumn{2}{c}{\textbf{ServiceNow}} &&
  \multicolumn{2}{c}{\textbf{GitLab}} &&
  \multicolumn{2}{c}{\textbf{ERPNext}} &&
  \multicolumn{2}{c}{\textbf{Overall}} \\
\textbf{Agent} && \multicolumn{2}{c}{(330)} &&
  \multicolumn{2}{c}{(192)} &&
  \multicolumn{2}{c}{(207)} &&
  \multicolumn{2}{c}{(729)} \\
\cmidrule(lr){3-4}
\cmidrule(lr){6-7}
\cmidrule(lr){9-10}
\cmidrule(lr){12-13}
&& \textit{SR~(\%)} & \textit{Cost~(\$)} &&
  \textit{SR~(\%)} & \textit{Cost~(\$)} &&
  \textit{SR~(\%)} & \textit{Cost~(\$)} &&
  \textit{SR~(\%)} & \textit{Cost~(\$)} \\
\midrule
\rowcolor{gray!10}\multicolumn{13}{c}{\textit{Gemma-4-31B-it}} \\
Tool-use &&
    9.4 & \textbf{0.005} &&
    39.9 & \textbf{0.012} &&
    30.4 & \textbf{0.007} &&
    23.4 & \textbf{0.007} \\
Web &&
    20.3 & 0.148 &&
    35.6 & 0.235 &&
    22.7 & 0.247 &&
    25.0 & 0.199 \\
Terminal &&
    \textbf{48.5} & 0.024 &&
    \textbf{52.1} & 0.016 &&
    \textbf{36.7} & 0.091 &&
    \textbf{46.1} & 0.041 \\
\midrule
\rowcolor{gray!10}\multicolumn{13}{c}{\textit{Qwen-3.6-27B}} \\
Tool-use &&
    10.9 & \textbf{0.024} &&
    50.5 & \textbf{0.037} &&
    48.3 & \textbf{0.030} &&
    31.9 & \textbf{0.029} \\
Web &&
    20.6 & 0.199 &&
    \textbf{59.6} & 0.109 &&
    34.3 & 0.319 &&
    34.8 & 0.209 \\
Terminal &&
    \textbf{49.4} & 0.065 &&
    \textbf{61.7} & 0.040 &&
    \textbf{56.0} & 0.063 &&
    \textbf{54.5} & 0.058 \\
\midrule
\rowcolor{gray!10}\multicolumn{13}{c}{\textit{Gemini 3.1 Pro}} \\
Tool-use &&
    14.2 & \textbf{0.10} &&
    48.9 & 0.15 &&
    62.8 & \textbf{0.07} &&
    37.1 & 0.11 \\
Web &&
    62.1 & 0.68 &&
    \textbf{84.6} & 0.22 &&
    65.2 & 1.13 &&
    68.9 & 0.69 \\
Terminal &&
    \textbf{78.5} & \textbf{0.10} &&
    79.8 & \textbf{0.06} &&
    \textbf{73.9} & 0.10 &&
    \textbf{77.5} & \textbf{0.09} \\
\bottomrule
\end{tabular}
\caption{\textbf{Results with open-weight models.}
Success rate (SR, $\uparrow$) and average cost per task ($\downarrow$) for tool-use, web, and terminal agents using Gemma-4-31B-it and Qwen-3.6-27B, alongside Gemini~3.1~Pro (the most cost-efficient frontier model in Table~\ref{tab:results-main}) as a reference point.
Overall results are weighted by the number of tasks in each platform.
Costs for the two open-weight models are reported to three decimal places, since they are an order of magnitude smaller than frontier-model costs and two decimals would collapse distinct values; the Gemini~3.1~Pro reference rows keep the two-decimal precision of Table~\ref{tab:results-main}.
Bold indicates the best SR and lowest cost within each platform--model group, along with any score within one standard error of the best.}
\label{tab:results-open-weight}
\end{table}

The results in Table~\ref{tab:results-open-weight} show that the advantage of terminal interaction is not restricted to frontier models.
Terminal agents obtain the highest success rate in every platform--model combination.
With Gemma-4-31B-it, the terminal agent achieves an overall success rate of 46.1\%, compared with 25.0\% for the web agent and 23.4\% for the tool-use agent.
With Qwen-3.6-27B, terminal performance increases to 54.5\%, compared with 34.8\% for web and 31.9\% for tool-use agents.

Although the open-weight models achieve lower absolute performance than the frontier models in Table~\ref{tab:results-main}, the relative ordering of the interaction paradigms remains consistent.
Moreover, terminal agents remain substantially more efficient than web agents: they are approximately $4.9\times$ cheaper with Gemma and $3.6\times$ cheaper with Qwen on average.

The tool-use results are more nuanced.
Tool-use agents outperform web agents in three of the six platform--model comparisons, suggesting that curated tools can benefit smaller models by reducing the need to discover API endpoints and construct requests from scratch.
Tool-use agents are also the least expensive in every comparison.
However, these benefits do not overcome the limitations of the off-the-shelf MCP servers evaluated here: terminal agents still achieve substantially higher overall success by retaining flexible access to the underlying platform APIs.

The three paradigms also do not degrade uniformly as model scale decreases. Relative to Gemini~3.1~Pro, overall success rate for web agents drops 43.9 points with Gemma and 34.1 points with Qwen (68.9\% $\to$ 25.0\% and 34.8\%), the steepest decline of the three. Terminal agents degrade less severely, dropping 31.4 points with Gemma and 23.0 points with Qwen (77.5\% $\to$ 46.1\% and 54.5\%). Tool-use agents degrade the least, by only 13.7 points with Gemma and 5.2 points with Qwen (37.1\% $\to$ 23.4\% and 31.9\%), consistent with curated tool registries lowering the burden of endpoint and payload discovery for weaker models.

Taken together, these results suggest that terminal agents do not succeed merely because frontier models are unusually capable coding agents; direct programmatic interaction remains effective even with substantially smaller open-weight backbones.

\subsection{Parametric knowledge vs.\ documentation}\label{sec:docs}

We investigate whether terminal agents benefit from access to official platform documentation or whether they can rely on their parametric knowledge and direct API interaction. Results are presented in Table~\ref{tab:results-docs}.          
                    
\begin{table}[htbp]
\vspace{-5pt}
\centering
\small
\setlength{\tabcolsep}{6pt}
\begin{tabular}{l c cc c cc c cc c cc}
\toprule
&&\multicolumn{2}{c}{\textbf{ServiceNow}} &&
  \multicolumn{2}{c}{\textbf{GitLab}} &&
  \multicolumn{2}{c}{\textbf{ERPNext}} &&
  \multicolumn{2}{c}{\textbf{Overall}} \\
\textbf{Agent} && \multicolumn{2}{c}{(330)} &&
  \multicolumn{2}{c}{(192)} &&
  \multicolumn{2}{c}{(207)} &&
  \multicolumn{2}{c}{(729)} \\
\cmidrule(lr){3-4}
\cmidrule(lr){6-7}
\cmidrule(lr){9-10}
\cmidrule(lr){12-13}
&& \textit{SR~(\%)} & \textit{Cost~(\$)} &&
  \textit{SR~(\%)} & \textit{Cost~(\$)} &&
  \textit{SR~(\%)} & \textit{Cost~(\$)} &&
  \textit{SR~(\%)} & \textit{Cost~(\$)} \\
\midrule
\rowcolor{gray!10}\multicolumn{13}{c}{\textit{Claude Sonnet 4.6}} \\
No Docs            &&
    \textbf{73.6} & \textbf{0.78} &&
    \textbf{76.5} & \textbf{0.28} &&
    67.6 & \textbf{0.46} &&
    \textbf{72.7} & \textbf{0.56} \\
With Docs            &&
    67.3 & 1.10 &&
    \textbf{79.1} & 0.48 &&
    \textbf{72.3} & 0.47 &&
    \textbf{71.8} & 0.76 \\
\midrule
\rowcolor{gray!10}\multicolumn{13}{c}{\textit{Claude Opus 4.6}} \\
No Docs            &&
    \textbf{79.1} & 1.94 &&
    \textbf{80.2} & \textbf{0.50} &&
    \textbf{76.8} & \textbf{0.72} &&
    \textbf{78.7} & 1.22 \\
With Docs            &&
    \textbf{81.2} & \textbf{1.60} &&
    \textbf{78.7} & 0.71 &&
    \textbf{76.3} & 0.99 &&
    \textbf{79.2} & \textbf{1.19} \\
\bottomrule
\end{tabular}
\vspace{-5pt}
\caption{\small{\textbf{Effect of documentation access on terminal agents.} Success rate and cost per task for terminal agents with and without access to official documentation. Bold indicates the best SR and lowest cost within each platform--model group, along with any score within one standard error of the best.}}
\label{tab:results-docs}
\vspace{-5pt}
\end{table}

Overall, documentation does not provide a clear benefit: both configurations achieve comparable success rates across both models we evaluated. With Sonnet 4.6, the agent without documentation access is slightly better overall (72.7\% vs.\ 71.8\%), while with Opus 4.6 the results are nearly identical (78.7\% vs.\ 79.2\%). This suggests that agents can often operate effectively without external documentation, either because relevant knowledge is internalized or because the APIs are sufficiently discoverable at inference time.

The aggregate numbers, however, mask divergent effects across platforms. On ServiceNow, documentation hurts with Sonnet ($-$6.3\%) and adds cost across both models, as agents spend a significant fraction of their tool call budget retrieving and reading documentation rather than executing the task. On ERPNext, documentation helps with Sonnet (+4.7\%) at virtually no additional cost, indicating that the agent can integrate the retrieved information efficiently. On GitLab, documentation has no effect on accuracy but roughly doubles the cost, suggesting the agent reads documentation without benefiting from it. We analyze why documentation helps on some platforms and hurts on others in Section~\ref{sec:docs-impact}.

This platform-level split should be interpreted with a caveat: ServiceNow and GitLab are widely used platforms with extensive public documentation, so it is plausible that our frontier model backbones already encode substantial knowledge of their APIs from pretraining, which would attenuate any incremental benefit from supplying documentation at inference time. ERPNext is comparatively less common, and it is precisely on ERPNext that documentation provides a clear benefit, consistent with this explanation. Our ablation cannot fully separate documentation structure (Section~\ref{sec:docs-impact}) from parametric familiarity with the platform as competing explanations for the ServiceNow and GitLab results, and we view this as a limitation of the current comparison rather than evidence that documentation is broadly unhelpful.

\subsection{How do terminal agents leverage documentation?}
\label{sec:docs-impact}

Table~\ref{tab:results-docs} shows that documentation can either help or hinder terminal agents. To better understand this effect, we compare runs from Claude Sonnet 4.6 with and without access to the official product documentation for both ServiceNow and ERPNext. We find that documentation can be either beneficial or detrimental depending on how well its structure matches the agent's needs.

ServiceNow's documentation is primarily written for users performing tasks through the web UI, and its API pages are \emph{reference-oriented}: they catalog the full API surface without indicating which endpoints are most appropriate for common operations. Agents reading these pages adopted unnecessarily complex strategies and spent a large fraction of their tool call budget on retrieval rather than task execution (Appendix~\ref{sec:doc-detrimental-case-study}). ERPNext's documentation is \emph{task-oriented}: each page describes a specific entity, its fields, and how to create or manage it, mapping directly to the API calls agents need to construct. This allowed agents to quickly find critical information, such as non-obvious doctype names, that cannot be discovered through schema exploration alone (Appendix~\ref{sec:doc-case-study}).

These findings suggest that documentation must be structured for its consumer. Human-oriented reference documentation can actively mislead agents, while concise, task-oriented content provides effective guidance. This is consistent with GitLab's well-documented, widely-known REST API yielding minimal skill accumulation in Section~\ref{sec:skills-discussion}, and motivates the use of \textit{Skills}, structured procedures written by agents themselves or by humans, as a complement to static documentation~\citep{li2026skillsbench}.

\subsection{Single agent vs.\ multi-agent systems}\label{sec:mas}

We compare a single terminal agent against a simple multi-agent system based on the planner-executor design~\citep{erdogan2025planandact}. Results are shown in Table~\ref{tab:results-mas}.

\begin{table}[ht]
\centering
\small
\setlength{\tabcolsep}{5pt}
\begin{tabular}{l c cc c cc c cc c cc}
\toprule
&&\multicolumn{2}{c}{\textbf{ServiceNow}} &&
  \multicolumn{2}{c}{\textbf{GitLab}} &&
  \multicolumn{2}{c}{\textbf{ERPNext}} &&
  \multicolumn{2}{c}{\textbf{Overall}} \\
\textbf{Agent} && \multicolumn{2}{c}{(330)} &&
  \multicolumn{2}{c}{(192)} &&
  \multicolumn{2}{c}{(207)} &&
  \multicolumn{2}{c}{(729)} \\
\cmidrule(lr){3-4}
\cmidrule(lr){6-7}
\cmidrule(lr){9-10}
\cmidrule(lr){12-13}
&& \textit{SR~(\%)} & \textit{Cost~(\$)} &&
  \textit{SR~(\%)} & \textit{Cost~(\$)} &&
  \textit{SR~(\%)} & \textit{Cost~(\$)} &&
  \textit{SR~(\%)} & \textit{Cost~(\$)} \\
\midrule
\rowcolor{gray!10}\multicolumn{13}{c}{\textit{Claude Sonnet 4.6}} \\
Single Agent            &&
    \textbf{73.6} & \textbf{0.78} &&
    76.5 & \textbf{0.28} &&
    67.6 & \textbf{0.46} &&
    72.7 & \textbf{0.56} \\
Planner/Executor            &&
    \textbf{71.5} & 1.13 &&
    \textbf{80.8} & 0.36 &&
    \textbf{74.4} & 0.68 &&
    \textbf{74.8} & 0.80 \\
\midrule
\rowcolor{gray!10}\multicolumn{13}{c}{\textit{Claude Opus 4.6}} \\
Single Agent            &&
    \textbf{79.1} & 1.94 &&
    \textbf{80.2} & \textbf{0.50} &&
    76.8 & \textbf{0.72} &&
    \textbf{78.7} & \textbf{1.22} \\
Planner/Executor            &&
    \textbf{79.7} & \textbf{1.68} &&
    76.1 & 0.63 &&
    \textbf{79.7} & 1.72 &&
    \textbf{78.8} & 1.41 \\
\bottomrule
\end{tabular}
\caption{\textbf{Single agent vs.\ planner-executor multi-agent system.} Success rate and cost per task for a single terminal agent compared to a planner-executor design where both agents share the same model and tool access. Bold indicates the best SR and lowest cost within each platform--model group, along with any score within one standard error of the best.}
\label{tab:results-mas}
\end{table}

With Sonnet 4.6, the planner-executor system achieves a slightly higher overall success rate (74.8\% vs.\ 72.7\%) but at 43\% higher cost (\$0.80 vs.\ \$0.56 per task).
With Opus 4.6, the gap narrows: both configurations achieve near-identical overall accuracy (78.7\% vs.\ 78.8\%), while the single agent remains cheaper (\$1.22 vs.\ \$1.41).
Overall, the planner-executor design provides a modest benefit on weaker models but the single-agent configuration remains competitive across all platforms and is consistently more cost-efficient.

\subsubsection{When does multi-agent orchestration help?}
\label{sec:mas-discussion}

The gains from the planner-executor design are not uniform across platforms or difficulty levels.
With Sonnet 4.6, the largest improvement appears on ERPNext (+6.8pp), where tasks often involve multi-step workflows with dependencies between records.
On ServiceNow, the single agent is slightly better (73.6\% vs.\ 71.5\%), and with Opus 4.6 the planner-executor actually regresses on GitLab (76.1\% vs.\ 80.2\%). This indicates that the planning overhead can be counterproductive when tasks are straightforward enough that direct execution suffices. Designing a process where the agent decides to invoke a planning agent before proceeding might give the main agent the flexibility to bypass this step if deemed unnecessary.

A closer look at ERPNext reveals that the planner-executor system disproportionately helps on harder tasks: of the 24 tasks where it improved over the single agent, 13 come from templates labelled hard (Appendix~\ref{sec:task-creation}), meaning their goals chain several dependent subtasks. These tasks typically require creating multiple linked records with non-obvious field configurations.
In these scenarios, the planner spends a substantial fraction of its budget exploring the system by querying DocType schemas, checking existing records, and verifying field semantics before producing a structured execution plan. This exploratory phase can surface information that a single agent, interleaving exploration with execution, is more likely to overlook. For instance, by querying a DocType schema rather than a record's current value, the planner may discover that a field's human-readable label contradicts its programmatic name, resolving an ambiguity before the executor encounters it (see example in Appendix~\ref{sec:mas-case-study}).

However, the benefit diminishes as model capability increases.
With Opus 4.6, the stronger model appears to internalize the planning step, and the overall accuracy gap disappears almost entirely. This suggests that explicit multi-agent orchestration primarily compensates for limitations in the underlying model's ability to reason over complex task structures, rather than providing an architectural advantage that scales with model capability.

\section{Results by Task Category}\label{sec:per-category}

Table~\ref{tab:results-main} reports success rate and cost per platform, which is the right granularity for the paradigm comparison but might be too coarse to say \emph{which} tasks each interface suits. This section breaks the same runs down by task category, so that the scope condition stated in the Limitations section and in Section~\ref{sec:cli-limitations} can be read off the data rather than asserted. No new agent runs are involved: these are re-analyses of the trajectories behind Table~\ref{tab:results-main}.

\paragraph{Conventions.} ServiceNow and GitLab use the categories of Appendix~\ref{sec:task-types} verbatim. We regroup ERPNext's tasks by operation (create, update, delete/cancel/disable, and composite tasks that chain dependent steps). Success rates and tokens are unweighted means over the four backbones of Table~\ref{tab:results-main}, so each model contributes equally to every category, and we report tokens rather than dollar cost because they are comparable across models and providers without carrying a pricing snapshot. Category sizes range from 2 to 126 tasks, so we test the terminal-versus-web difference with a paired exact McNemar test over the identical task set, pooling discordant pairs across backbones. \textbf{Differences without a significant $p$-value should be read as ties, not as small wins.}

\begin{table}[ht]
\centering
\small
\setlength{\tabcolsep}{4pt}
\begin{tabular}{l r r ccc ccc}
\toprule
&&& \multicolumn{3}{c}{\textit{Success rate (\%)}} & \multicolumn{3}{c}{\textit{Tokens per task (K)}} \\
\cmidrule(lr){4-6}\cmidrule(lr){7-9}
\textbf{Category} & \textbf{n} & \textit{p} & \textit{Tool-use} & \textit{Web} & \textit{Term.} & \textit{Tool-use} & \textit{Web} & \textit{Term.} \\
\midrule
\rowcolor{gray!10}\multicolumn{9}{c}{\textit{ServiceNow}} \\
Form Creation             & 50 & \textbf{$<$0.001} & 2.0  & 67.0 & \textbf{89.0} & 59  & 513 & 42 \\
Dashboard                 & 40 & 0.302 & 11.9 & 36.2 & 39.1 & 107 & 599 & 186 \\
List Filtering            & 60 & 0.072 & 20.8 & 70.0 & 76.7 & 50  & 397 & 34 \\
List Sorting              & 60 & \textbf{0.003} & 35.8 & 62.5 & \textbf{75.4} & 87  & 174 & 249 \\
Knowledge Base            & 10 & 1.000 & 87.5 & 90.0 & 87.5 & 391 & 139 & 14 \\
Navigation                & 10 & 0.065 & 15.0 & 52.5 & 70.0 & 24  & 236 & 11 \\
Service Catalog           & 90 & \textbf{0.005} & 0.0  & 90.6 & \textbf{96.1} & 77  & 306 & 53 \\
Impersonation             & 10 & \textbf{$<$0.001} & 0.0  & \textbf{95.0} & 0.0  & 30  & 176 & 28 \\
\midrule
\rowcolor{gray!10}\multicolumn{9}{c}{\textit{GitLab}} \\
Create                    & 49 & 0.388 & 44.1 & 85.6 & 82.0 & 91  & 112 & 38 \\
Read / List               & 55 & \textbf{0.033} & 58.9 & \textbf{74.9} & 67.6 & 142 & 72  & 34 \\
Update                    & 28 & 0.424 & 49.1 & 89.3 & 85.7 & 65  & 134 & 44 \\
Social                    & 49 & \textbf{$<$0.001} & 37.2 & \textbf{81.0} & 71.1 & 74  & 70  & 34 \\
Misc                      & 2  & 1.000 & 12.5 & 50.0 & 50.0 & 34  & 393 & 46 \\
Composite                 & 9  & 1.000 & 0.0  & 47.2 & 47.2 & 98  & 159 & 104 \\
\midrule
\rowcolor{gray!10}\multicolumn{9}{c}{\textit{ERPNext} (regrouped by operation)} \\
Create                    & 126 & 0.727 & 64.9 & 63.5 & 64.1 & 44 & 511 & 79 \\
Update                    & 34  & 0.134 & 61.8 & 92.6 & 86.8 & 19 & 205 & 57 \\
Delete, cancel or disable & 22  & 0.289 & 50.0 & 92.0 & 96.6 & 9  & 74  & 22 \\
Composite                 & 25  & \textbf{0.009} & 61.0 & 57.0 & \textbf{71.0} & 30 & 402 & 54 \\
\bottomrule
\end{tabular}
\caption{\textbf{Success rate and token cost by task category.} Unweighted means over the four backbones of Table~\ref{tab:results-main}. $p$ is a paired exact McNemar test of terminal against web over the tasks solved by exactly one of the two, and bold marks the categories it separates at $p < 0.05$, together with the paradigm that wins there; every other row is a tie. Tokens are input plus output per task, in thousands, and include failed attempts and tasks that hit the turn limit, so a paradigm that fails expensively is charged for it.}
\label{tab:per-category}
\end{table}

\paragraph{Most categories are ties; the terminal agent wins where operations must be composed.} The test separates terminal from web agents in 7 of the 18 categories. The four terminal wins share a mechanism: the requested effect has to be assembled rather than clicked. ServiceNow form creation, catalog requests and saved list views each require getting fields, variants or query state right, and ERPNext composites chain dependent documents. Two of the three web wins are the cases Section~\ref{sec:cli-limitations} predicts, impersonation and GitLab's one-click social actions. The third is GitLab's read/list category, where listing records is a tie, and the deficit comes from tasks whose answer must be \emph{computed} (e.g. ranking contributors by commit count and counting a user's commits on a given day). GitLab renders both aggregates, so the web agent reads a number the terminal agent reconstructs from repository history, as with ServiceNow dashboards. Per backbone, the decided categories point the same way for all four models, while in the tie categories the sign of the difference changes from model to model. Sorting is the only task category where conclusions differ by backbone: three backbones favor the terminal agent by $12$ to $32$ points and Opus~4.6 favors the web agent by $22$.

\paragraph{Cost separates almost everywhere, and does not follow the same pattern.} The web agent spends $1.5$ to $20.8\times$ more tokens per task than the terminal agent in every category but one, including the categories where it is more accurate. The exception is ServiceNow list sorting, where the terminal agent spends more ($249$K versus $174$K) while scoring $12.9$ points higher, consistent with agents re-querying the list to check the resulting order. Tool-use agents are the cheapest paradigm in most categories, with only two expensive categories: on ServiceNow knowledge-base search ($391$K) and GitLab read/list tasks ($142$K) the catalog exposes no tool that returns the requested information, so the agent spends its budget calling tools that cannot answer the question.

\paragraph{What this scopes.} The API-first default applies when the effect a task asks for is a change to platform state that the API exposes, and it weakens when that effect lives in the browser session, in information the interface renders rather than stores, or behind a single UI affordance that no endpoint mirrors. Within that scope the two paradigms are at parity on most categories, the terminal agent's accuracy advantage is concentrated in operations that must be composed rather than clicked, and its cost advantage holds nearly everywhere. The practical reading is the one Section~\ref{sec:hybrid-discussion} draws: default to API-first access, and escalate to a browser for the classes above rather than for tasks in general.

\section{Unintended State Changes}\label{sec:side-effects}

Our validators check that the requested state exists at the end of a trajectory, not that nothing else changed, so an agent that reaches that state while also creating a second record, or while writing to something the task never mentioned, is scored as a success. We therefore audit every state-changing request in the 2{,}911 terminal and 2{,}914 tool-use trajectories behind Table~\ref{tab:results-main}. This is a request-level audit of traces rather than a diff of the environment: it sees what each agent issued and what the platform returned instead of what the environment finally held. \textbf{The web agent is not comparable} on this measure, since a click that submits a form is indistinguishable in the trace from a click that opens a menu.

The terminal agent writes far more often, on 73.0\% of tasks against 47.4\% and at 3.30 writes per task against 1.23, but that volume is retry rather than exploration: 61.8\% of its tasks send two or more writes to one endpoint against 20.2\% for tool-use, and the platform refuses 31\% of its attempts against 2\%. The tool-use agent's restraint is not caution. On 12 ServiceNow tasks its most frequent write was creating a script include, which runs arbitrary server-side code with the credential's privileges. A narrow catalog therefore constrains what an agent can be asked \emph{through the tools}, not what the credential permits, and the workaround it invites is harder to audit than the request it replaces.

Duplicates are rare, and our validators cannot see them. We look for them on the 200 ServiceNow form-creation tasks, which each ask for a single record: a trajectory that ends up with two records in the same table has created one the task did not ask for. This happens on 2 of the 200 tasks for the terminal agent and on none for the tool-use agent, too few to read as a rate. One of the two is still scored as a success, because the agent created \texttt{PRB0040003} and \texttt{PRB0040004} for one requested problem and the validator only asks whether a matching problem exists.

Writes the task never asked for are rare as well. We count them on the information-only tasks, where the goal is to retrieve something and any write is therefore unrequested: 9 of 465 such tasks for the terminal agent and 2 of 466 for the tool-use agent. The clearest case is a GitLab task that asks only to read an RSS feed token, where three of the four backbones tried to rotate the token instead.

Quantifying either properly needs validators that grade the end state and assert the \emph{absence} of changes the task did not ask for, as AutomationBench does with negative assertions~\citep{shepard2026automationbench} and AppWorld with checks on unintended state~\citep{trivedi2024appworld}. Our tasks are also single-session and single-platform, so we expect side effects to matter more on harder benchmarks than they do here. An unintended write in our setting affects one record. The same write in an environment with business-rule cascades~\citep{gupta2026world} or policy blockers~\citep{malay2026enterpriseopsgymenvironmentsevaluationsstateful} can trigger downstream changes that our tasks have no way to produce.

\section{Consistency Across Task Instances}\label{sec:consistency}

Success rate answers how often an agent completes a task drawn from the benchmark. Deployment asks a stricter question: whether the agent handles \emph{every} instance of a recurring task, since an automation that succeeds four times in five still has to be supervised.

Our ServiceNow benchmark is built from 33 task types with exactly 10 seeds each, where a seed varies the records, filters, and field values while keeping the operation fixed (Appendix~\ref{sec:task-creation}). Treating a type's 10 seeds as 10 draws from its difficulty distribution, pass\textasciicircum{}$k$ is the probability that all $k$ drawn instances succeed and pass@$k$ that at least one does. With $s$ successes out of $n$ seeds both are exact under sampling without replacement, $\binom{s}{k}/\binom{n}{k}$ for pass\textasciicircum{}$k$, and we bootstrap over task types for confidence intervals. This is \emph{not} the pass\textasciicircum{}$k$ of $\tau$-bench~\citep{yao2024taubench}, which repeats the same instance and isolates run-to-run stochasticity. Ours varies the instance, so it answers whether the agent handles every incident it is given rather than whether it is deterministic; our estimate of run-to-run variation comes from the repeated runs in Appendix~\ref{sec:reproducibility}.

\begin{table}[ht]
\centering
\small
\setlength{\tabcolsep}{5pt}
\begin{tabular}{l cccc c ccc}
\toprule
& \multicolumn{4}{c}{pass\textasciicircum{}$k$ \textit{(all $k$ instances succeed)}} && \multicolumn{3}{c}{pass@$k$} \\
\cmidrule(lr){2-5}\cmidrule(lr){7-9}
\textbf{Agent} & $k{=}1$ & $k{=}3$ & $k{=}5$ & $k{=}10$ && $k{=}3$ & $k{=}5$ & $k{=}10$ \\
\midrule
\rowcolor{gray!10}\multicolumn{9}{c}{\textit{Claude Sonnet 4.6}} \\
Terminal & 73.3 & 43.9 & 27.6 & 9.1  && 94.8 & 96.7 & 97.0 \\
Web      & 72.7 & 44.5 & 30.9 & 15.6 && 95.7 & 99.3 & 100.0 \\
Tool-use & 11.8 & 2.0  & 0.7  & 0.0  && 25.1 & 32.8 & 42.4 \\
\midrule
\rowcolor{gray!10}\multicolumn{9}{c}{\textit{Claude Opus 4.6}} \\
Terminal & 78.7 & 59.6 & 50.3 & 40.6 && 94.1 & 96.4 & 96.9 \\
Web      & 77.9 & 60.9 & 53.6 & 48.4 && 92.8 & 96.2 & 100.0 \\
Tool-use & 16.1 & 3.4  & 1.8  & 0.0  && 33.4 & 43.5 & 54.5 \\
\midrule
\rowcolor{gray!10}\multicolumn{9}{c}{\textit{GPT-5.4 Thinking (Medium)}} \\
Terminal & 77.0 & 65.5 & 57.6 & 45.5 && 84.5 & 84.8 & 84.8 \\
Web      & 69.4 & 46.4 & 33.5 & 15.2 && 86.3 & 87.8 & 87.9 \\
Tool-use & 18.5 & 8.4  & 3.9  & 0.0  && 27.7 & 30.4 & 33.3 \\
\midrule
\rowcolor{gray!10}\multicolumn{9}{c}{\textit{Gemini 3.1 Pro}} \\
Terminal & 78.5 & 57.9 & 46.8 & 33.3 && 94.2 & 96.2 & 97.0 \\
Web      & 62.1 & 31.4 & 20.5 & 12.1 && 90.5 & 97.3 & 100.0 \\
Tool-use & 14.2 & 3.2  & 1.6  & 0.0  && 28.1 & 34.8 & 42.4 \\
\bottomrule
\end{tabular}
\caption{\textbf{Consistency across instances of a task type}, averaged over the ServiceNow task types. pass\textasciicircum{}$k$ is the probability that all $k$ instances drawn from a task type succeed, pass@$k$ that at least one does, and $k{=}1$ is the success rate. Instances vary records and field values within a fixed operation, so these numbers measure robustness across instances rather than run-to-run determinism.}
\label{tab:consistency}
\end{table}

No paradigm is reliable at the level its success rate suggests. Every paradigm degrades sharply from $k{=}1$ to $k{=}10$: a terminal agent at 78.7\% success handles all ten instances of only 40.6\% of task types with Opus~4.6, and 9.1\% with Sonnet~4.6. The tool-use agent's deficit is sharper here than on success rate, reaching pass\textasciicircum{}10 $=0.0$ for all four backbones while its pass@10 of 33 to 55\% shows many of those types are within reach on a good attempt. The terminal--web comparison, at parity on success rate, becomes backbone-dependent: the terminal agent is far more consistent with GPT-5.4 (45.5 versus 15.2) and Gemini~3.1~Pro (33.3 versus 12.1) and less consistent with the two Claude models, so we claim no consistency advantage for either interface. Within the terminal agent's own results, consistency survives where the operation is a well-specified write, with catalog requests and form creation holding 69.4\% and 55.0\% at $k{=}10$, and collapses to 0.0\% on every category that requires reading or computing a displayed value.

\section{Task Generation}\label{sec:task-creation}

\subsection{Task Design}

For ServiceNow and GitLab, we adapt tasks from prior benchmarks~\citep{drouin2024workarena, zhou2024webarena} but redesign the evaluation pipeline (see below).
For ERPNext, we construct a new benchmark from scratch: professional linguists designed the seed data populating each instance, wrote the natural-language goals, and authored the SQL validation queries.
Every goal template and validation query was manually verified against sample data in a live instance.

The ERPNext templates are additionally labelled by difficulty, which we define compositionally: a \emph{hard} template chains several dependent subtasks into a single goal, typically creating multiple linked records whose identifiers or field values depend on records created earlier in the chain, whereas the remaining templates target a single operation. This mirrors the compositional levels of WorkArena++~\citep{boisvert2024workarena}, whose L2 and L3 tasks are likewise built by chaining atomic tasks.

Each task is instantiated from a template with specific seed data, yielding 330 ServiceNow tasks across 33 templates, 192 GitLab tasks, and 207 ERPNext tasks (see Appendix~\ref{sec:task-types} for the full breakdown).
Tasks span several categories: record creation, retrieval, update, deletion, filtering, sorting, navigation, and multi-step composite workflows.

Two design choices distinguish our tasks from prior work.
First, we add a \emph{navigation} requirement to every task: agents are never given a start URL or a direct link to the relevant record.
Instead, they must determine which API endpoints or platform pages to interact with based solely on the natural-language goal.
Second, we systematically remove ambiguity from goal descriptions.
For example, a ServiceNow task originally phrased as ``sort the list by priority'' is rewritten to ``sort the \emph{incident} list by priority,'' since the agent has no browser context to disambiguate the target table.
Together with the rebuilt validators described below, these changes mean that our ServiceNow and GitLab success rates are not directly comparable to published WorkArena or WebArena results: the tasks are harder to locate, more precisely specified, and scored against platform state rather than against the original benchmarks' criteria. All comparisons we report are therefore between agent paradigms within our own evaluation pipeline.

\subsection{Evaluation}

The environment is fully reset before each task: the platform instance is restored to a fixed snapshot containing the seed data, ensuring that tasks are independent and cannot interfere with one another.

Validators check the platform state before and after the agent finishes, rather than inspecting the agent's actions or intermediate outputs. The check prior to the agent's trajectory is necessary to ensure that the task does not pass without having the agent perform any action.
For ServiceNow and GitLab, validators issue API calls (via Python \texttt{requests}) against the live instance to verify that the expected records were created, updated, or match the specified criteria.
For ERPNext, validators execute SQL queries directly against the database.
In both cases, the validator compares the post-execution state against the expected outcome defined in the task template.

For read-oriented tasks such as retrieving a list of records matching a filter, or navigating to a specific page, the validator parses the agent's final message to extract the returned answer or URL, then checks it against the ground truth. This ensures that even tasks without write side-effects are evaluated against the actual platform state rather than relying on string matching alone.

\clearpage

\section{Taxonomy of tool call outcomes} \label{sec:tool-call-outcome-taxonomy}

% Requires: \usepackage{booktabs}, \usepackage{colortbl}, \usepackage{xcolor}
\begin{table*}[ht]
\centering
\fontsize{8}{9}\selectfont
\begin{tabular}{l p{5cm} p{7.5cm}}
\toprule
\textbf{Category} & \textbf{Description} & \textbf{Example output} \\
\midrule
\rowcolor{gray!10}
\multicolumn{3}{c}{\textit{Success categories}} \\
\addlinespace[3pt]
Success
  & Valid JSON response with expected data.
  & \texttt{\{"result": \{"sys\_id": "a1b2...", "number": "INC001"\}\}} \\
\addlinespace[2pt]
Success (trunc.)
  & Valid response cut off by output length limit.
  & \texttt{\{"result": [\{"sys\_id": ...\} ...] [OUTPUT TRUNCATED]} \\
\addlinespace[2pt]
Non-JSON success
  & Output piped through \texttt{jq}, \texttt{python3}, or \texttt{awk}; non-JSON but correct.
  & \texttt{number: INC0000039 state: 6} \\
\midrule
\rowcolor{gray!10}
\multicolumn{3}{c}{\textit{Failure categories}} \\
\addlinespace[3pt]
API error
  & \parbox[t]{5cm}{\raggedright
Server returned a JSON error response\\
(invalid payload, bad field, wrong HTTP method).}
  & \texttt{\{"error": \{"detail": "The payload is not valid JSON."\}, "status": "failure"\}} \\
\addlinespace[2pt]
Shell error
  & Non-zero exit code or syntax error. Often from unescaped characters in \texttt{-d} payloads with \texttt{eval}.
  & \texttt{/bin/sh: Syntax error: "\}" unexpected} \\
\addlinespace[2pt]
Empty response
  & No output returned. Typically from wrong HTTP method (e.g., \texttt{PUT} instead of \texttt{PATCH}).
  & \texttt{[no output]} \\
\addlinespace[2pt]
Curl error
  & Curl failed before reaching the server. Multi-line \texttt{-d} with \texttt{eval} splits the command.
  & \texttt{curl: (2) no URL specified} \\
\addlinespace[2pt]
JSON parse error
  & Output piped to a JSON parser that failed on non-JSON upstream output.
  & \texttt{parse error: Invalid literal at line 1} \\
\addlinespace[2pt]
Python error
  & Inline \texttt{python3 -c} post-processing raised an exception on unexpected input.
  & \texttt{json.decoder.JSONDecodeError: Expecting value} \\
\addlinespace[2pt]
Timeout
  & Command exceeded the 30\,s execution limit on slow or unbounded queries.
  & \texttt{[error] Command timed out after 30s.} \\
\addlinespace[2pt]
HTML redirect
  & Auth redirect instead of JSON; agent bypassed provided headers.
  & \texttt{<html>...<meta http-equiv="refresh" ...>} \\
\bottomrule
\end{tabular}

\caption{Classification of tool call outcomes observed in evaluation traces. The agent interacts with web platforms through \texttt{curl} commands, whose outputs are categorized into success or failure types based on the response content.}
% \vspace{-0.3cm}
\label{tab:tool-call-outcome-taxonomy}
\end{table*}

\clearpage

\section{Full Results}\label{sec:full-results}

\subsection{Full Results: Comparing Agent Interaction Paradigms}
\begin{table}[ht]
\centering
\small
\setlength{\tabcolsep}{2.5pt}
\begin{tabular}{l c cccc c cccc c cccc c cccc}
\toprule
&&\multicolumn{4}{c}{\textbf{ServiceNow}} &&
  \multicolumn{4}{c}{\textbf{GitLab}} &&
  \multicolumn{4}{c}{\textbf{ERPNext}} &&
  \multicolumn{4}{c}{\textbf{Overall}} \\
\textbf{Agent} && \multicolumn{4}{c}{(330)} &&
  \multicolumn{4}{c}{(192)} &&
  \multicolumn{4}{c}{(207)} &&
  \multicolumn{4}{c}{(729)} \\
\cmidrule(lr){3-6}
\cmidrule(lr){8-11}
\cmidrule(lr){13-16}
\cmidrule(lr){18-21}
&& \textit{SR} & \textit{Tools} & \textit{Time} & \textit{Cost} &&
  \textit{SR} & \textit{Tools} & \textit{Time} & \textit{Cost} &&
  \textit{SR} & \textit{Tools} & \textit{Time} & \textit{Cost} &&
  \textit{SR} & \textit{Tools} & \textit{Time} & \textit{Cost} \\
\midrule
\rowcolor{gray!10}\multicolumn{21}{c}{\textit{Claude Sonnet 4.6}} \\
Tool-use            &&
    11.5 & \textbf{4.3} & \textbf{35} & \textbf{0.76} &&
    45.2 & \textbf{4.0} & \textbf{22} & 0.48 &&
    55.6 & \textbf{5.7} & \textbf{17} & \textbf{0.14} &&
    32.9 & \textbf{4.6} & \textbf{26} & \textbf{0.51} \\
Web            &&
    \textbf{72.4} & 26.0 & 153 & 4.49 &&
    \textbf{82.9} & 10.4 & 55 & 0.88 &&
    61.8 & 19.9 & 138 & 3.63 &&
    \textbf{72.2} & 20.2 & 123 & 3.29 \\
Terminal            &&
    \textbf{73.6} & 10.5 & 67 & 0.78 &&
    76.5 & 8.8 & 42 & \textbf{0.28} &&
    \textbf{67.6} & 13.0 & 42 & 0.46 &&
    \textbf{72.7} & 10.8 & 53 & 0.56\\
\midrule
\rowcolor{gray!10}\multicolumn{21}{c}{\textit{Claude Opus 4.6}} \\
Tool-use            &&
    16.1 & \textbf{3.1} & \textbf{31} & \textbf{0.66} &&
    46.8 & \textbf{4.2} & \textbf{25} & 0.90 &&
    68.9 & \textbf{5.9} & \textbf{25} & \textbf{0.17} &&
    39.2 & \textbf{4.2} & \textbf{28} & \textbf{0.58} \\
Web            &&
    \textbf{77.6} & 19.8 & 141 & 4.21 &&
    \textbf{81.9} & 8.1 & 51 & 0.85 &&
    \textbf{81.6} & 24.4 & 128 & 6.49 &&
    \textbf{79.9} & 18.0 & 114 & 3.97 \\
Terminal            &&
    \textbf{79.1} & 13.9 & 103 & 1.94 &&
    \textbf{80.2} & 9.9 & 50 & \textbf{0.50} &&
    76.8 & 14.5 & 59 & 0.72 &&
    \textbf{78.7} & 13.0 & 77 & 1.22 \\
\midrule
% \rowcolor{gray!10}\multicolumn{21}{c}{\textit{GPT-5.2 (Medium)}} \\
% Tool-use            &&
%     16.7 & \textbf{2.9} & \textbf{31} & \textbf{0.10} &&
%     47.9 & \textbf{5.9} & \textbf{59} & 0.25 &&
%     68.6 & \textbf{14.1} & \textbf{62} & \textbf{0.19} &&
%     39.7 & \textbf{6.9} & \textbf{47} & \textbf{0.16} \\
% Web            &&
%     \textbf{75.8} & 21.2 & 176 & 0.66 &&
%     \textbf{72.9} & 7.8 & 80 & \textbf{0.18} &&
%     \textbf{76.3} & 30.4 & 168 & 1.33 &&
%     \textbf{75.2} & 20.3 & 148 & 0.73 \\
% Terminal            &&
%     \textbf{77.3} & 15.7 & 89 & 0.22 &&
%     \textbf{71.8} & 14.0 & 84 & \textbf{0.17} &&
%     70.5 & 29.0 & 151 & 0.48 &&
%     \textbf{73.9} & 19.0 & 105 & 0.28 \\
% \midrule
\rowcolor{gray!10}\multicolumn{21}{c}{\textit{GPT-5.4 Thinking (Medium)}} \\
Tool-use            &&
     18.5 & \textbf{3.1} & \textbf{51} & \textbf{0.14} &&
     47.9 & \textbf{6.4} & \textbf{57} & 0.40 &&
     62.8 & \textbf{11.2} & \textbf{66} & \textbf{0.21} &&
     38.8 & \textbf{6.3} & \textbf{57} & 0.23 \\
Web            &&
     69.4 & 11.9 & 132 & 0.54 &&
     \textbf{81.4} & 6.5 & 80 & 0.17 &&
     \textbf{72.5} & 12.1 & 128 & 0.51 &&
     \textbf{73.4} & 10.5 & 117 & 0.43 \\
Terminal            &&
     \textbf{77.0} & 9.8 & 95 & 0.20 &&
     71.3 & 9.6 & 78 & \textbf{0.13} &&
     \textbf{70.0} & 18.0 & 115 & 0.24 &&
     \textbf{73.5} & 12.1 & 96 & \textbf{0.19} \\
\midrule
\rowcolor{gray!10}\multicolumn{21}{c}{\textit{Gemini 3.1 Pro}} \\
Tool-use            &&
    14.2 & \textbf{2.2} & \textbf{30} & \textbf{0.10} &&
    48.9 & \textbf{2.5} & \textbf{28} & 0.15 &&
    62.8 & \textbf{6.3} & \textbf{30} & \textbf{0.07} &&
    37.1 & \textbf{3.4} & \textbf{29} & 0.11 \\
Web            &&
    62.1 & 13.3 & 109 & 0.68 &&
    \textbf{84.6} & 7.8 & 68 & 0.22 &&
    65.2 & 17.6 & 136 & 1.13 &&
    68.9 & 13.1 & 106 & 0.69 \\
Terminal            &&
    \textbf{78.5} & 6.0 & 55 & \textbf{0.10} &&
    79.8 & 7.5 & 53 & \textbf{0.06} &&
    \textbf{73.9} & 9.8 & 53 & 0.10 &&
    \textbf{77.5} & 7.5 & 54 & \textbf{0.09} \\
\bottomrule
\end{tabular}
\caption{\textbf{Experiment results of StarShell on enterprise benchmarks.} Success rate (SR~$\uparrow$),
average tool calls (Tools), wall-clock time in seconds (Time), and cost per task in USD (Cost~$\downarrow$)
are reported for tool-use, web, and terminal agents on 3 platforms for 4 backbone LLMs. Bold indicates the best value for each metric within a model group, along with any SR within one standard error of the best.}
\label{tab:results-full}
\end{table}

\subsection{Full Results: Comparing Terminal Agents With and Without Access to Documentation}
\begin{table}[ht]
\centering
\small
\setlength{\tabcolsep}{2.5pt}
\begin{tabular}{l c cccc c cccc c cccc c cccc}
\toprule
&&\multicolumn{4}{c}{\textbf{ServiceNow}} &&
  \multicolumn{4}{c}{\textbf{GitLab}} &&
  \multicolumn{4}{c}{\textbf{ERPNext}} &&
  \multicolumn{4}{c}{\textbf{Overall}} \\
\textbf{Agent} && \multicolumn{4}{c}{(330)} &&
  \multicolumn{4}{c}{(192)} &&
  \multicolumn{4}{c}{(207)} &&
  \multicolumn{4}{c}{(729)} \\
\cmidrule(lr){3-6}
\cmidrule(lr){8-11}
\cmidrule(lr){13-16}
\cmidrule(lr){18-21}
&& \textit{SR} & \textit{Tools} & \textit{Time} & \textit{Cost} &&
  \textit{SR} & \textit{Tools} & \textit{Time} & \textit{Cost} &&
  \textit{SR} & \textit{Tools} & \textit{Time} & \textit{Cost} &&
  \textit{SR} & \textit{Tools} & \textit{Time} & \textit{Cost} \\
\midrule
\rowcolor{gray!10}\multicolumn{21}{c}{\textit{Claude Sonnet 4.6}} \\
No Docs        &&
    \textbf{73.6} & \textbf{10.5} & \textbf{67} & \textbf{0.78} &&
    \textbf{76.5} & \textbf{8.8} & \textbf{42} & \textbf{0.28} &&
    67.6 & \textbf{13.0} & \textbf{42} & \textbf{0.46} &&
    \textbf{72.7} & \textbf{10.8} & \textbf{53} & \textbf{0.56} \\
With Docs      &&
    67.3 & 18.0 & 94 & 1.10 &&
    \textbf{79.1} & 12.2 & 45 & 0.48 &&
    \textbf{72.3} & 15.3 & 52 & 0.47 &&
    \textbf{71.8} & 15.7 & 69 & 0.76 \\
\midrule
\rowcolor{gray!10}\multicolumn{21}{c}{\textit{Claude Opus 4.6}} \\
No Docs        &&
    \textbf{79.1} & \textbf{13.9} & \textbf{103} & 1.94 &&
    \textbf{80.2} & \textbf{9.9} & \textbf{50} & \textbf{0.50} &&
    \textbf{76.8} & \textbf{14.5} & \textbf{59} & \textbf{0.72} &&
    \textbf{78.7} & \textbf{13.0} & \textbf{77} & 1.22 \\
With Docs      &&
    \textbf{81.2} & 16.5 & 105 & \textbf{1.60} &&
    \textbf{78.7} & 13.0 & 59 & 0.71 &&
    \textbf{76.3} & 18.0 & 75 & 0.99 &&
    \textbf{79.2} & 16.0 & 84 & \textbf{1.19} \\
\bottomrule
\end{tabular}
\caption{\textbf{Effect of documentation access on terminal agents.} Success rate (SR~$\uparrow$),
average tool calls (Tools), wall-clock time in seconds (Time), and cost per task in USD (Cost~$\downarrow$)
are reported for terminal agents with and without access to official platform documentation. Bold indicates the best value for each metric within a model group, along with any SR within one standard error of the best.}
\label{tab:results-docs-full}
\end{table}

\clearpage

\subsection{Full Results: Agents Equipped with Skills}

\begin{table}[ht]
\vspace{-5pt}
\centering
\small
\setlength{\tabcolsep}{2.5pt}
\begin{tabular}{l c cccc c cccc c cccc c cccc}
\toprule
&&\multicolumn{4}{c}{\textbf{ServiceNow}} &&
  \multicolumn{4}{c}{\textbf{GitLab}} &&
  \multicolumn{4}{c}{\textbf{ERPNext}} &&
  \multicolumn{4}{c}{\textbf{Overall}} \\
\textbf{Agent} && \multicolumn{4}{c}{(330)} &&
  \multicolumn{4}{c}{(192)} &&
  \multicolumn{4}{c}{(207)} &&
  \multicolumn{4}{c}{(729)} \\
\cmidrule(lr){3-6}
\cmidrule(lr){8-11}
\cmidrule(lr){13-16}
\cmidrule(lr){18-21}
&& \textit{SR} & \textit{Tools} & \textit{Time} & \textit{Cost} &&
  \textit{SR} & \textit{Tools} & \textit{Time} & \textit{Cost} &&
  \textit{SR} & \textit{Tools} & \textit{Time} & \textit{Cost} &&
  \textit{SR} & \textit{Tools} & \textit{Time} & \textit{Cost} \\
\midrule
\rowcolor{gray!10}\multicolumn{21}{c}{\textit{Claude Sonnet 4.6}} \\
No Skills        &&
    73.6 & 10.5 & \textbf{67} & 0.78 &&
    \textbf{76.5} & \textbf{8.8} & \textbf{42} & \textbf{0.28} &&
    67.6 & 13.0 & \textbf{42} & 0.46 &&
    72.7 & 10.8 & \textbf{53} & 0.56 \\
With Skills      &&
    \textbf{77.0} & \textbf{8.7} & \textbf{67} & \textbf{0.44} &&
    \textbf{78.1} & 9.9 & 50 & 0.31 &&
    \textbf{72.9} & \textbf{11.7} & 62 & \textbf{0.35} &&
    \textbf{76.1} & \textbf{9.9} & 61 & \textbf{0.38} \\
\bottomrule
\end{tabular}
\caption{\textbf{Effect of the ability to create and reuse skills on terminal agents.} Success rate (SR~$\uparrow$),
average tool calls (Tools), wall-clock time in seconds (Time), and cost per task in USD (Cost~$\downarrow$)
are reported for terminal agents with and without access to self-generated skills. Bold indicates the best value for each metric within a model group, along with any SR within one standard error of the best.}
\label{tab:results-skill-full}
\end{table}

\subsection{Full Results: Single Agent vs Multi-Agent System}

\begin{table}[ht]
\vspace{-5pt}
\centering
\small
\setlength{\tabcolsep}{1.8pt}
\begin{tabular}{l c cccc c cccc c cccc c cccc}
\toprule
&&\multicolumn{4}{c}{\textbf{ServiceNow}} &&
  \multicolumn{4}{c}{\textbf{GitLab}} &&
  \multicolumn{4}{c}{\textbf{ERPNext}} &&
  \multicolumn{4}{c}{\textbf{Overall}} \\
\textbf{Agent} && \multicolumn{4}{c}{(330)} &&
  \multicolumn{4}{c}{(192)} &&
  \multicolumn{4}{c}{(207)} &&
  \multicolumn{4}{c}{(729)} \\
\cmidrule(lr){3-6}
\cmidrule(lr){8-11}
\cmidrule(lr){13-16}
\cmidrule(lr){18-21}
&& \textit{SR} & \textit{Tools} & \textit{Time} & \textit{Cost} &&
  \textit{SR} & \textit{Tools} & \textit{Time} & \textit{Cost} &&
  \textit{SR} & \textit{Tools} & \textit{Time} & \textit{Cost} &&
  \textit{SR} & \textit{Tools} & \textit{Time} & \textit{Cost} \\
\midrule
\rowcolor{gray!10}\multicolumn{21}{c}{\textit{Claude Sonnet 4.6}} \\
Single Agent       &&
    \textbf{73.6} & \textbf{10.5} & \textbf{67} & \textbf{0.78} &&
    76.5 & \textbf{8.8} & \textbf{42} & \textbf{0.28} &&
    67.6 & \textbf{13.0} & \textbf{42} & \textbf{0.46} &&
    72.7 & \textbf{10.8} & \textbf{53} & \textbf{0.56}\\
Planner/Executor   &&
    \textbf{71.5} & 16.6 & 103 & 1.13 &&
    \textbf{80.8} & 10.9 & 54 & 0.36 &&
    \textbf{74.4} & 20.9 & 109 & 0.68 &&
    \textbf{74.8} & 16.3 & 92 & 0.80 \\
\midrule
\rowcolor{gray!10}\multicolumn{21}{c}{\textit{Claude Opus 4.6}} \\
Single Agent       &&
    \textbf{79.1} & \textbf{13.9} & \textbf{103} & 1.94 &&
    \textbf{80.2} & \textbf{9.9} & \textbf{50} & \textbf{0.50} &&
    76.8 & \textbf{14.5} & \textbf{59} & \textbf{0.72} &&
    \textbf{78.7} & \textbf{13.0} & \textbf{77} & \textbf{1.22} \\
Planner/Executor   &&
    \textbf{79.7} & 16.5 & 123 & \textbf{1.68} &&
    76.1 & 13.5 & 78 & 0.63 &&
    \textbf{79.7} & 27.6 & 150 & 1.72 &&
    \textbf{78.8} & 18.9 & 119 & 1.41 \\
\bottomrule
\end{tabular}
\caption{\textbf{Single agent vs.\ planner-executor multi-agent system.} Success rate (SR~$\uparrow$),
average tool calls (Tools), wall-clock time in seconds (Time), and cost per task in USD (Cost~$\downarrow$)
are reported for a single terminal agent compared to a planner-executor design where both agents share the same model and tool access. Bold indicates the best value for each metric within a model group, along with any SR within one standard error of the best.}
\label{tab:results-mas-full}
\end{table}

\subsection{Full results: Hybrid Agents}

\begin{table}[ht]
\vspace{-5pt}
\centering
\small
\setlength{\tabcolsep}{2.5pt}
\begin{tabular}{l c cccc c cccc c cccc c cccc}
\toprule
&&\multicolumn{4}{c}{\textbf{ServiceNow}} &&
  \multicolumn{4}{c}{\textbf{GitLab}} &&
  \multicolumn{4}{c}{\textbf{ERPNext}} &&
  \multicolumn{4}{c}{\textbf{Overall}} \\
\textbf{Agent} && \multicolumn{4}{c}{(330)} &&
  \multicolumn{4}{c}{(192)} &&
  \multicolumn{4}{c}{(207)} &&
  \multicolumn{4}{c}{(729)} \\
\cmidrule(lr){3-6}
\cmidrule(lr){8-11}
\cmidrule(lr){13-16}
\cmidrule(lr){18-21}
&& \textit{SR} & \textit{Tools} & \textit{Time} & \textit{Cost} &&
  \textit{SR} & \textit{Tools} & \textit{Time} & \textit{Cost} &&
  \textit{SR} & \textit{Tools} & \textit{Time} & \textit{Cost} &&
  \textit{SR} & \textit{Tools} & \textit{Time} & \textit{Cost} \\
\midrule
\rowcolor{gray!10}\multicolumn{21}{c}{\textit{Claude Sonnet 4.6}} \\
Web            &&
    \textbf{72.4} & 26.0 & 153 & 4.49 &&
    \textbf{82.9} & 10.4 & 55 & 0.88 &&
    61.8 & 19.9 & 138 & 3.63 &&
    \textbf{72.2} & 20.2 & 123 & 3.29 \\
Terminal            &&
    \textbf{73.6} & \textbf{10.5} & \textbf{67} & \textbf{0.78} &&
    76.5 & \textbf{8.8} & \textbf{42} & \textbf{0.28} &&
    \textbf{67.6} & \textbf{13.0} & \textbf{42} & \textbf{0.46} &&
    \textbf{72.7} & \textbf{10.8} & \textbf{53} & \textbf{0.56}\\
Hybrid   &&
    \textbf{72.1} & 21.0 & 127 & 3.02 &&
    77.7 & 10.5 & 49 & 0.64 &&
    \textbf{66.7} & 16.5 & 89 & 1.44 &&
    \textbf{72.0} & 17.0 & 96 & 1.94 \\
\midrule
\rowcolor{gray!10}\multicolumn{21}{c}{\textit{Claude Opus 4.6}} \\
Web            &&
    77.6 & 19.8 & 141 & 4.21 &&
    \textbf{81.9} & \textbf{8.1} & 51 & 0.85 &&
    \textbf{81.6} & 24.4 & 128 & 6.49 &&
    79.9 & 18.0 & 114 & 3.97 \\
Terminal       &&
    79.1 & \textbf{13.9} & 103 & \textbf{1.94} &&
    \textbf{80.2} & 9.9 & \textbf{50} & \textbf{0.50} &&
    76.8 & \textbf{14.5} & \textbf{59} & \textbf{0.72} &&
    78.7 & \textbf{13.0} & \textbf{77} & \textbf{1.22} \\
Hybrid   &&
    \textbf{83.0} & 15.0 & \textbf{100} & 2.57 &&
    \textbf{80.3} & 10.7 & 55 & 0.94 &&
    \textbf{79.7} & 19.8 & 98 & 2.84 &&
    \textbf{81.4} & 15.2 & 88 & 2.22 \\
\bottomrule
\end{tabular}
\caption{\textbf{Hybrid agent comparison.} Success rate (SR~$\uparrow$),
average tool calls (Tools), wall-clock time in seconds (Time), and cost per task in USD (Cost~$\downarrow$)
are reported for a hybrid agent using the Playwright MCP and a terminal, compared against a simple terminal agent and an agent equipped solely with the Playwright MCP tools. Bold indicates the best value for each metric within a model group, along with any SR within one standard error of the best.}
\label{tab:results-hybrid-full}
\end{table}

\clearpage

\section{Task Types per Benchmark}\label{sec:task-types}

\begin{table*}[h]
\centering
\fontsize{8}{9}\selectfont
\setlength{\tabcolsep}{3pt}
\begin{minipage}[t]{0.49\textwidth}
\centering
\subsection{ServiceNow}
\caption{ServiceNow tasks}
\label{tab:servicenow-task-overview}
\begin{tabular}{llr}
\toprule
\textbf{Category} & \textbf{Task Name} & \textbf{\# Tasks} \\
\midrule
\multirow{5}{*}{Form Creation}
  & create-change-request & 10 \\
  & create-hardware-asset & 10 \\
  & create-incident & 10 \\
  & create-problem & 10 \\
  & create-user & 10 \\
\midrule
\multirow{4}{*}{Dashboard}
  & single-chart-value-retrieval & 10 \\
  & single-chart-min-max-retrieval & 10 \\
  & multi-chart-value-retrieval & 10 \\
  & multi-chart-min-max-retrieval & 10 \\
\midrule
\multirow{6}{*}{List Filtering}
  & filter-asset-list & 10 \\
  & filter-change-request-list & 10 \\
  & filter-hardware-list & 10 \\
  & filter-incident-list & 10 \\
  & filter-service-catalog-item-list & 10 \\
  & filter-user-list & 10 \\
\midrule
\multirow{6}{*}{List Sorting}
  & sort-asset-list & 10 \\
  & sort-change-request-list & 10 \\
  & sort-hardware-list & 10 \\
  & sort-incident-list & 10 \\
  & sort-service-catalog-item-list & 10 \\
  & sort-user-list & 10 \\
\midrule
Knowledge Base & knowledge-base-search & 10 \\
\midrule
Navigation & all-menu & 10 \\
\midrule
\multirow{9}{*}{Service Catalog}
  & order-apple-mac-book-pro15 & 10 \\
  & order-apple-watch & 10 \\
  & order-developer-laptop & 10 \\
  & order-development-laptop-p-c & 10 \\
  & order-ipad-mini & 10 \\
  & order-ipad-pro & 10 \\
  & order-loaner-laptop & 10 \\
  & order-sales-laptop & 10 \\
  & order-standard-laptop & 10 \\
\midrule
Impersonation & impersonation & 10 \\
\midrule
Total & & 330 \\
\bottomrule
\end{tabular}
\end{minipage}
\hfill
\begin{minipage}[t]{0.49\textwidth}
\centering
\subsection{GitLab}

\caption{GitLab tasks}
\label{tab:gitlab-task-overview}

\begin{tabular}{llr}
\toprule
\textbf{Category} & \textbf{Task Name} & \textbf{\# Tasks} \\
\midrule
\multirow{5}{*}{Create}
  & create-group & 5 \\
  & create-issue & 13 \\
  & create-milestone & 5 \\
  & create-pr & 6 \\
  & create-repo & 20 \\
\midrule
\multirow{6}{*}{Read / List}
  & list-issues & 12 \\
  & list-pr & 2 \\
  & list-repos & 6 \\
  & list-info & 12 \\
  & show-command & 5 \\
  & count-commit & 18 \\
\midrule
\multirow{6}{*}{Update}
  & update-issue & 5 \\
  & update-license & 5 \\
  & update-multiple-issues-single-repo & 3 \\
  & update-profile & 5 \\
  & update-repo & 5 \\
  & update-status & 5 \\
\midrule
\multirow{6}{*}{Social}
  & follow-user & 5 \\
  & fork-repo & 6 \\
  & invite-user & 15 \\
  & open-issue & 10 \\
  & post-message & 8 \\
  & star-repo & 5 \\
\midrule
\multirow{2}{*}{Misc}
  & get-rss-token & 1 \\
  & get-todo & 1 \\
\midrule
\multirow{3}{*}{Composite}
  & add-followed-users-to-new-repo & 3 \\
  & get-and-update-info & 3 \\
  & star-fork-create-milestone-issues & 3 \\
\midrule
Total & & 192 \\
\bottomrule
\end{tabular}
\end{minipage}

\end{table*}

\newpage

{\fontsize{8}{9}\selectfont

\subsection{ERPNext}
\vspace{0.3cm}
\begin{longtable}{llr}
\caption{ERPNext Tasks}
\small
\label{tab:erpnext-task-overview} \\
\toprule
\textbf{Category} & \textbf{Task Name} & \textbf{\# Tasks} \\
\midrule
\endfirsthead
\toprule
\textbf{Category} & \textbf{Task Name} & \textbf{\# Tasks} \\
\midrule
\endhead
\midrule
\multicolumn{3}{r}{\textit{Continued on next page}} \\
\endfoot
\bottomrule
\endlastfoot
% ---- ERPNext (207) ----
\multirow{6}{*}{User Management}
  & assign-role & 2 \\
  & create-user-email-first-last & 2 \\
  & create-user-email-first-last-username-phone & 3 \\
  & create-user-group-with-new-users & 3 \\
  & create-user-with-email-account-dependency & 3 \\
  & disable-user-account & 2 \\
\midrule
\multirow{6}{*}{Purchase Orders}
  & add-comment-to-draft-purchase-orders & 1 \\
  & add-date-confirmation-number-to-purchase-order & 2 \\
  & create-purchase-order-then-manage-supplier & 2 \\
  & create-purchase-order-with-discount & 2 \\
  & create-purchase-order-without-discount & 2 \\
  & update-target-warehouse-on-suppliers-purchase-orders & 2 \\
\midrule
\multirow{4}{*}{Supplier Management}
  & create-supplier-name-email-phone & 2 \\
  & create-supplier-name-group-currency-tax-id & 2 \\
  & delete-supplier & 3 \\
  & tag-supplier & 3 \\
\midrule
\multirow{6}{*}{Customer Management}
  & create-customer-name-account-manager-group & 3 \\
  & create-customer-name-type-address & 3 \\
  & create-customer-then-create-sales-invoice & 3 \\
  & create-customer-then-create-sales-order & 3 \\
  & create-customer-with-account-manager-group-tax-dep. & 3 \\
  & create-customer-with-acct-mgr-role-profile-group-tax-dep. & 3 \\
\midrule
\multirow{10}{*}{Sales}
  & assign-sales-invoice & 2 \\
  & create-sales-invoice & 3 \\
  & create-sales-invoice-with-custom-tax-rate & 3 \\
  & create-sales-invoice-with-customer-item-account-dep. & 3 \\
  & create-sales-invoice-with-cust-item-terr-pricelist-acct-dep. & 3 \\
  & create-sales-invoice-with-date-cost-center & 3 \\
  & create-sales-order & 3 \\
  & create-sales-order-put-on-hold & 3 \\
  & favorite-sales-invoice & 1 \\
  & tag-sales-invoice & 2 \\
\midrule
\multirow{11}{*}{Inventory}
  & create-stock-entry & 2 \\
  & create-stock-entry-with-letterhead & 2 \\
  & create-stock-item-id-group & 3 \\
  & create-stock-item-name-id-group-description & 3 \\
  & create-stock-item-name-id-group-desc-shelf-life-warranty & 3 \\
  & create-stock-item-with-name-desc-then-create-po & 2 \\
  & create-stock-item-without-name-desc-then-create-po & 2 \\
  & delete-stock-item & 2 \\
  & disable-stock-item & 2 \\
  & edit-stock-item-warranty-weight & 2 \\
  & edit-stock-item-warranty-weight-material-req-type-tag & 2 \\
\midrule
\multirow{5}{*}{Employees}
  & create-employee-full-address & 3 \\
  & create-employee-name-gender-dob-join-date-salary-phone & 3 \\
  & create-employee-name-gender-dob-status-join-date & 3 \\
  & create-employee-with-emergency-contact & 3 \\
  & update-employee-department & 2 \\
\midrule
\multirow{10}{*}{Accounting}
  & assign-journal-entry & 2 \\
  & assign-payment-entry & 2 \\
  & cancel-journal-entry & 3 \\
  & cancel-payment-entry & 3 \\
  & create-journal-entry & 2 \\
  & create-journal-entry-with-account-dependency & 2 \\
  & create-payment-entry-with-cost-center-cheque-date-ref & 2 \\
  & create-payment-entry-with-cust-acct-project-dep. & 3 \\
  & create-payment-entry-with-mop-acct-supplier-cc-dep. & 3 \\
  & create-payment-entry-without-cost-center & 2 \\
\midrule
\multirow{17}{*}{Projects \& Tasks}
  & assign-task & 3 \\
  & change-project-status & 2 \\
  & create-project-with-customer-full-address & 3 \\
  & create-project-with-name-company & 3 \\
  & create-project-with-name-company-priority-note & 3 \\
  & create-project-with-name-company-priority-note-dept-cost & 3 \\
  & create-project-with-type-dept-cost-center-dep. & 3 \\
  & create-project-with-type-dept-customer-cc-dep. & 3 \\
  & create-project-with-type-dept-dep. & 3 \\
  & create-task-with-project-type-dept-dep. & 3 \\
  & create-task-with-project-type-parent-dept-dep. & 3 \\
  & create-task-with-subject-project & 2 \\
  & create-task-with-subject-project-priority-dept & 3 \\
  & create-task-with-subject-project-priority-dept-comment & 3 \\
  & delete-project & 2 \\
  & delete-task & 3 \\
  & update-task-status & 2 \\
\midrule
\multirow{7}{*}{Composite}
  & assign-journal-entry-then-update-task-status & 2 \\
  & assign-purchase-order & 2 \\
  & cancel-purchase-order & 2 \\
  & create-project-then-cancel-payment-entry & 3 \\
  & create-sales-invoice-then-assign-task & 3 \\
  & create-task-then-assign-payment-entry & 2 \\
  & create-user-then-cancel-journal-entry & 3 \\
\midrule
Total & & 207 \\
\end{longtable}
}

\newpage

\section{MCP Servers}\label{sec:mcp-servers}

\subsection{ServiceNow}

% Requires: \usepackage{longtable, booktabs, multirow}
{
\vspace{-0.3cm}

\fontsize{8}{9}\selectfont
\begin{longtable}{lllr}
\caption{Tools advertised by the ServiceNow MCP server (83 tools). The server's \texttt{tool\_packages.yaml} lists 93 tools in its default \texttt{full} package, and its startup log reports that figure, but 11 of the 93 have no implementation and are silently omitted from the \texttt{ListTools} response; \texttt{list\_tool\_packages} is added by the server itself. We report the 83 the model can actually invoke.}
\vspace{-0.3cm}
\label{tab:servicenow-mcp-tools} \\
\toprule
\textbf{Category} & \textbf{Tool Name} & & \textbf{\# Tools} \\
\midrule
\endfirsthead
\toprule
\textbf{Category} & \textbf{Tool Name} & & \textbf{\# Tools} \\
\midrule
\endhead
\midrule
\multicolumn{4}{r}{\textit{Continued on next page}} \\
\endfoot
\bottomrule
\endlastfoot
% ============================================================
% ServiceNow MCP (83 tools advertised; echelon-ai-labs@0625060, MCP_TOOL_PACKAGE=full)
% ============================================================
\multirow{3}{*}{Incident Management}
  & create\_incident & resolve\_incident & \multirow{3}{*}{6}\\
  & update\_incident & list\_incidents &\\
  & add\_comment & get\_incident\_by\_number &\\
\midrule
\multirow{6}{*}{Service Catalog}
  & list\_catalog\_items & move\_catalog\_items & \multirow{6}{*}{11}\\
  & get\_catalog\_item & get\_optimization\_recommendations &\\
  & update\_catalog\_item & create\_catalog\_item\_variable &\\
  & list\_catalog\_categories & list\_catalog\_item\_variables &\\
  & create\_catalog\_category & update\_catalog\_item\_variable &\\
  & update\_catalog\_category &  &\\
\midrule
\multirow{4}{*}{Change Management}
  & create\_change\_request & add\_change\_task & \multirow{4}{*}{8}\\
  & update\_change\_request & submit\_change\_for\_approval &\\
  & list\_change\_requests & approve\_change &\\
  & get\_change\_request\_details & reject\_change &\\
\midrule
\multirow{8}{*}{Agile Management}
  & create\_story & list\_epics & \multirow{8}{*}{15}\\
  & update\_story & create\_scrum\_task &\\
  & list\_stories & update\_scrum\_task &\\
  & create\_story\_dependency & list\_scrum\_tasks &\\
  & delete\_story\_dependency & create\_project &\\
  & list\_story\_dependencies & update\_project &\\
  & create\_epic & list\_projects &\\
  & update\_epic &  &\\
\midrule
\multirow{6}{*}{Workflow Management}
  & list\_workflows & list\_workflow\_versions & \multirow{6}{*}{12}\\
  & create\_workflow & get\_workflow\_activities &\\
  & update\_workflow & add\_workflow\_activity &\\
  & activate\_workflow & update\_workflow\_activity &\\
  & deactivate\_workflow & delete\_workflow\_activity &\\
  & get\_workflow\_details & reorder\_workflow\_activities &\\
\midrule
\multirow{3}{*}{Script Includes}
  & list\_script\_includes & update\_script\_include & \multirow{3}{*}{5}\\
  & get\_script\_include & delete\_script\_include &\\
  & create\_script\_include &  &\\
\midrule
\multirow{4}{*}{Changeset Management}
  & list\_changesets & commit\_changeset & \multirow{4}{*}{7}\\
  & get\_changeset\_details & publish\_changeset &\\
  & create\_changeset & add\_file\_to\_changeset &\\
  & update\_changeset &  &\\
\midrule
\multirow{5}{*}{Knowledge Base}
  & create\_knowledge\_base & update\_article & \multirow{5}{*}{9}\\
  & list\_knowledge\_bases & publish\_article &\\
  & create\_category & list\_articles &\\
  & list\_categories & get\_article &\\
  & create\_article &  &\\
\midrule
\multirow{5}{*}{User Management}
  & create\_user & update\_group & \multirow{5}{*}{9}\\
  & update\_user & add\_group\_members &\\
  & get\_user & remove\_group\_members &\\
  & list\_users & list\_groups &\\
  & create\_group &  &\\
\midrule
Configuration
  & list\_tool\_packages &  & 1\\
\midrule
Total & & & 83 \\
\end{longtable}
}
\vspace{-0.6cm}

\subsection{ERPNext}

{
\vspace{-0.4cm}
\fontsize{8}{9}\selectfont
\begin{longtable}{lllr}
\caption{Overview of tools available in the ERPNext MCP server.}
\vspace{-0.3cm}
\label{tab:erpnext-mcp-tools} \\
\toprule
\textbf{Category} & \textbf{Tool Name} & & \textbf{\# Tools} \\
\midrule
\endfirsthead
\toprule
\textbf{Category} & \textbf{Tool Name} & & \textbf{\# Tools} \\
\midrule
\endhead
\midrule
\multicolumn{4}{r}{\textit{Continued on next page}} \\
\endfoot
\bottomrule
\endlastfoot
% ============================================================
% ERPNext MCP (7 tools)
% ============================================================
Authentication
  & authenticate\_erpnext & & 1\\
\midrule
\multirow{2}{*}{Document Management}
  & get\_documents & update\_document & \multirow{2}{*}{3}\\
  & create\_document & \\
\midrule
\multirow{1}{*}{Schema}
  & get\_doctypes & get\_doctype\_fields & 2\\
\midrule
Reporting
  & run\_report & & 1\\
\midrule
Total & & & 7 \\
\end{longtable}
}

\newpage

\subsection{GitLab}

{\fontsize{8}{9}\selectfont
\begin{longtable}{lllr}
\caption{Tools advertised by the GitLab MCP server (81 tools). The server additionally implements wiki, milestone and pipeline tool groups, which it exposes only when the \texttt{USE\_GITLAB\_WIKI}, \texttt{USE\_MILESTONE} and \texttt{USE\_PIPELINE} environment variables are set; we leave them at their defaults, so those 26 tools were unavailable to the agent and appear in no recorded trajectory.}

\label{tab:gitlab-mcp-tools} \\
\toprule
\textbf{Category} & \textbf{Tool Name} & & \textbf{\# Tools} \\
\midrule
\endfirsthead
\toprule
\textbf{Category} & \textbf{Tool Name} & & \textbf{\# Tools} \\
\midrule
\endhead
\midrule
\multicolumn{4}{r}{\textit{Continued on next page}} \\
\endfoot
\bottomrule
\endlastfoot
% ============================================================
% GitLab MCP (81 tools advertised; zereight@v2.0.30, wiki/milestone/pipeline groups disabled)
% ============================================================
\multirow{16}{*}{Merge Requests}
  & create\_merge\_request & update\_merge\_request\_note & \multirow{16}{*}{31}\\
  & get\_merge\_request & delete\_merge\_request\_note &\\
  & update\_merge\_request & create\_merge\_request\_thread &\\
  & merge\_merge\_request & resolve\_merge\_request\_thread &\\
  & list\_merge\_requests & mr\_discussions &\\
  & get\_merge\_request\_diffs & create\_merge\_request\_discussion\_note &\\
  & list\_merge\_request\_diffs & update\_merge\_request\_discussion\_note &\\
  & get\_merge\_request\_version & delete\_merge\_request\_discussion\_note &\\
  & list\_merge\_request\_versions & get\_draft\_note &\\
  & approve\_merge\_request & list\_draft\_notes &\\
  & unapprove\_merge\_request & create\_draft\_note &\\
  & get\_merge\_request\_approval\_state & update\_draft\_note &\\
  & create\_note & delete\_draft\_note &\\
  & create\_merge\_request\_note & publish\_draft\_note &\\
  & get\_merge\_request\_note & bulk\_publish\_draft\_notes &\\
  & get\_merge\_request\_notes &  &\\
\midrule
\multirow{7}{*}{Issues}
  & create\_issue & update\_issue\_note & \multirow{7}{*}{13}\\
  & list\_issues & list\_issue\_discussions &\\
  & my\_issues & list\_issue\_links &\\
  & get\_issue & get\_issue\_link &\\
  & update\_issue & create\_issue\_link &\\
  & delete\_issue & delete\_issue\_link &\\
  & create\_issue\_note &  &\\
\midrule
\multirow{4}{*}{Repositories}
  & search\_repositories & push\_files & \multirow{4}{*}{7}\\
  & create\_repository & fork\_repository &\\
  & get\_file\_contents & get\_repository\_tree &\\
  & create\_or\_update\_file &  &\\
\midrule
\multirow{3}{*}{Branches \& Commits}
  & create\_branch & get\_commit & \multirow{3}{*}{5}\\
  & get\_branch\_diffs & get\_commit\_diff &\\
  & list\_commits &  &\\
\midrule
\multirow{4}{*}{Projects}
  & get\_project & get\_namespace & \multirow{4}{*}{8}\\
  & list\_projects & verify\_namespace &\\
  & list\_project\_members & list\_group\_projects &\\
  & list\_namespaces & list\_group\_iterations &\\
\midrule
\multirow{3}{*}{Labels}
  & list\_labels & update\_label & \multirow{3}{*}{5}\\
  & get\_label & delete\_label &\\
  & create\_label &  &\\
\midrule
\multirow{4}{*}{Releases}
  & list\_releases & delete\_release & \multirow{4}{*}{7}\\
  & get\_release & create\_release\_evidence &\\
  & create\_release & download\_release\_asset &\\
  & update\_release &  &\\
\midrule
\multirow{3}{*}{Users \& Events}
  & get\_users & upload\_markdown & \multirow{3}{*}{5}\\
  & list\_events & download\_attachment &\\
  & get\_project\_events &  &\\
\midrule
Total & & & 81 \\
\end{longtable}
}

\newpage

\section{Playwright MCP Server}\label{sec:playwright-mcp-server}

% Requires: \usepackage{longtable, booktabs, multirow}
\begin{longtable}{llr}
\caption{Tools advertised by the Playwright MCP server\protect\footnote{\url{https://github.com/microsoft/playwright-mcp}, v0.0.68, the release current throughout our evaluation window.} in the configuration used by our web and hybrid agents (21 tools). The server additionally implements vision (coordinate-level mouse), PDF, and tracing tool groups, which are gated behind opt-in capability flags that we leave disabled; \texttt{browser\_install} is advertised but blocked by our harness, since browsers are pre-installed in the container. None of these tools were available to the agent, and none appear in any recorded trajectory.}
\label{tab:playwright-mcp-tools} \\
\toprule
\textbf{Category} & \textbf{Tool Name} & \textbf{\# Tools} \\
\midrule
\endfirsthead
\toprule
\textbf{Category} & \textbf{Tool Name} & \textbf{\# Tools} \\
\midrule
\endhead
\midrule
\multicolumn{3}{r}{\textit{Continued on next page}} \\
\endfoot
\bottomrule
\endlastfoot
\multirow{20}{*}{Core Automation}
  & browser\_click & \multirow{20}{*}{20}\\
  & browser\_close & \\
  & browser\_console\_messages & \\
  & browser\_drag & \\
  & browser\_evaluate & \\
  & browser\_file\_upload & \\
  & browser\_fill\_form & \\
  & browser\_handle\_dialog & \\
  & browser\_hover & \\
  & browser\_navigate & \\
  & browser\_navigate\_back & \\
  & browser\_network\_requests & \\
  & browser\_press\_key & \\
  & browser\_resize & \\
  & browser\_run\_code & \\
  & browser\_select\_option & \\
  & browser\_snapshot & \\
  & browser\_take\_screenshot & \\
  & browser\_type & \\
  & browser\_wait\_for & \\
\midrule
Tab Management
  & browser\_tabs & 1 \\
\midrule
\textbf{Total} & & \textbf{21} \\
\end{longtable}

\newpage

\section{Error Analysis Tooling}\label{sec:tooling}

\begin{figure}[h]
    \centering
    \includegraphics[width=\textwidth]{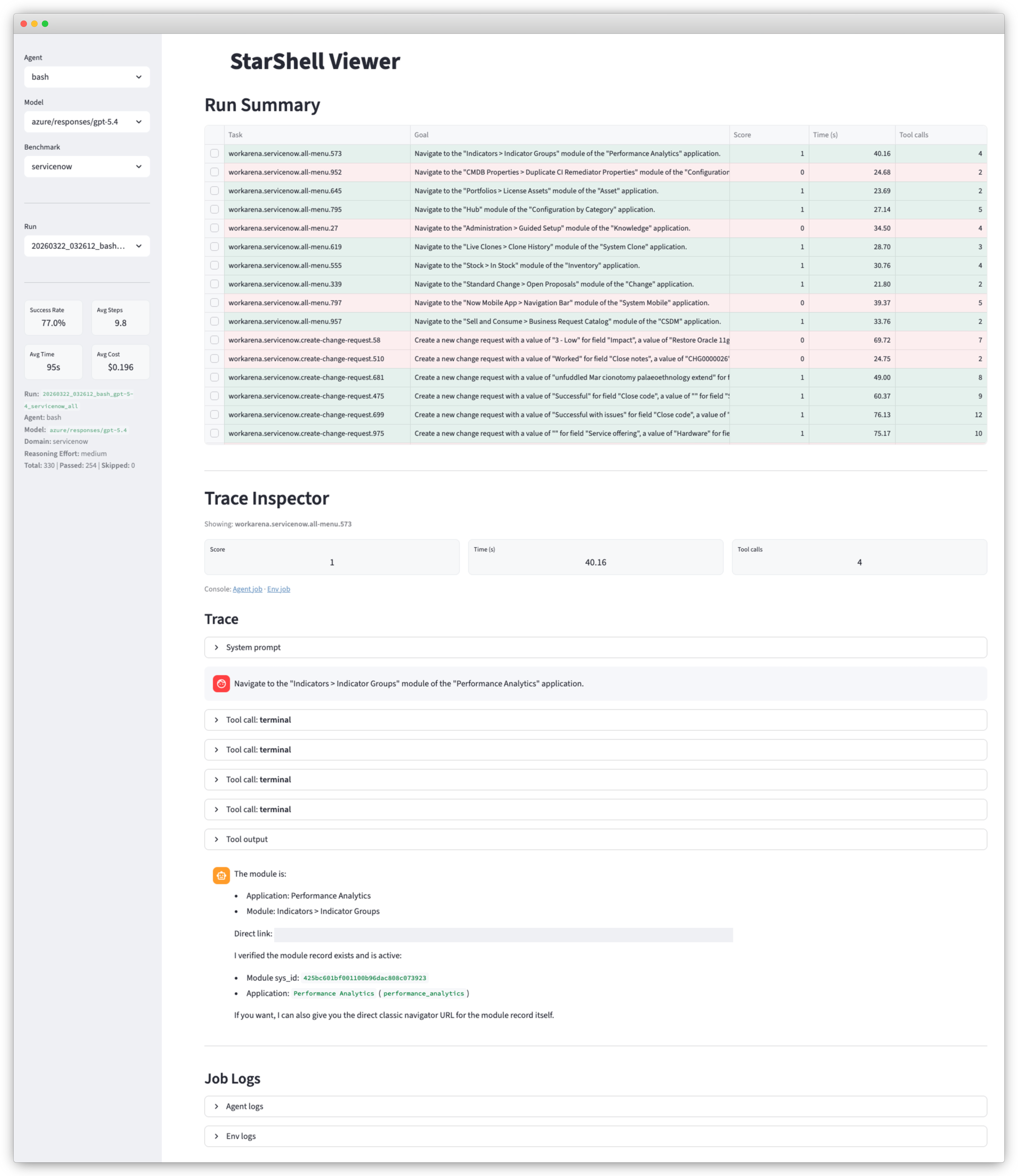}
    \caption{User interface used to inspect agent traces.}
    \label{fig:error-analysis-tooling}
\end{figure}

Figure~\ref{fig:error-analysis-tooling} depicts the webapp used to perform error analysis on agent traces. The interface includes high-level metrics such as success rate and number of tools used on average for a run, a table containing outcomes for each sample for a given benchmark, and a viewer to see the agent's reasoning when executing the task. We also provide the detailed logs from both the agent run and the environment.

\clearpage

\section{Prompts}\label{sec:prompts}

\subsection{Tool-use Agent Prompt for ServiceNow}

\begin{tcblisting}{
    colback=gray!5,
    colframe=black!60,
    title={Prompt for Tool-use Agent for ServiceNow},
    breakable,
    listing only
}
You are a ServiceNow assistant with a rich set of ServiceNow MCP tools. Use the available tools to interact with the ServiceNow instance directly - you do not need to make raw API calls or use curl.

## ServiceNow Instance

URL: <url>

## Available Tool Categories

You have many tools organized by function:

- **Incident Management** - create, update, comment, resolve, and list incidents
- **Service Catalog** - manage catalog items, categories, variables, and catalogs
- **Change Management** - create/update change requests, add tasks, submit for approval, approve/reject changes
- **Knowledge Base** - create/manage knowledge bases, categories, articles; publish articles
- **Agile / Project Management** - stories, epics, scrum tasks, projects, story dependencies
- **User & Group Management** - create/update users and groups, manage membership
- **Workflows** - list, create, update, delete workflows
- **Script Includes** - list, create, update, delete script includes
- **Changesets** - manage changesets, add files, commit, publish
- **UI Policies** - create UI policies and policy actions for catalog forms

## How to work

1. Use the appropriate MCP tool for each task. Do not try to work around the tools - they are your primary interface to ServiceNow.
2. When creating or updating records, confirm the result by reading back the record after the operation.
3. **When you complete a task**, always provide the user with a direct URL link to the relevant ServiceNow record so they can verify the result. For example: `<url>/now/nav/ui/classic/params/target/incident.do?sys_id=<sys_id>`.
4. If a tool returns an error, read the error message carefully and adjust your approach - do not retry the exact same call.
5. When listing records, use filters to narrow results rather than fetching everything.
\end{tcblisting}

\clearpage
\subsection{Tool-use Agent Prompt for GitLab}

\begin{tcblisting}{
    colback=gray!5,
    colframe=black!60,
    title={Prompt for Tool-use Agent for GitLab},
    breakable,
    listing only
}
You are a GitLab assistant with a rich set of GitLab MCP tools. Use the available tools to interact with the GitLab instance directly - you do not need to make raw API calls or use curl.

## GitLab Instance

URL: <url>

## Available Tool Categories

You have many tools organized by function:

- **Project Management** - create, update, list, and delete projects
- **Issue Tracking** - create, update, comment on, and close issues
- **Merge Requests** - create, review, approve, and merge MRs
- **CI/CD Pipelines** - trigger, monitor, and manage pipelines and jobs
- **Repository** - browse files, branches, tags, and commits
- **User & Group Management** - manage users, groups, and memberships
- **Wiki** - create and manage project wiki pages
- **Snippets** - create and manage code snippets

## How to work

1. Use the appropriate MCP tool for each task. Do not try to work around the tools - they are your primary interface to GitLab.
2. When creating or updating records, confirm the result by reading back the record after the operation.
3. **When you complete a task**, always provide the user with a direct URL link to the relevant GitLab record so they can verify the result. For example: `<url>/<namespace>/<project>/-/issues/<id>`.
4. If a tool returns an error, read the error message carefully and adjust your approach - do not retry the exact same call.
5. When listing records, use filters to narrow results rather than fetching everything.
- **When the user asks to list multiple items** (e.g. issues with a certain label, merge requests assigned to a user, projects a user contributed to), always provide a clickable GitLab URL that shows those results in the web UI. For example:
  - Issues with label `bug`: `<url>/<namespace>/<project>/-/issues?label_name[]=bug`
  - MRs assigned to a user: `<url>/<namespace>/<project>/-/merge_requests?assignee_username=<user>`
  - All MRs by an author: `<url>/dashboard/merge_requests?author_username=<user>`
  This lets the user click through and see the full, live list directly in GitLab.

\end{tcblisting}
\clearpage
\subsection{Tool-use Agent Prompt for ERPNext}

\begin{tcblisting}{
    colback=gray!5,
    colframe=black!60,
    title={Prompt for Tool-use Agent for ERPNext},
    breakable,
    listing only
}
You are a ERPNext assistant with a rich set of ERPNext MCP tools. Use the available tools to interact with the ERPNext instance directly - you do not need to make raw API calls or use curl.

## ERPNext Instance

URL: <url>

## Available Tool Categories

You have many tools organized by function:

- **Accounting** - manage invoices, journal entries, and chart of accounts
- **Selling** - quotations, sales orders, and customer management
- **Buying** - purchase orders, supplier quotations, suppliers
- **Stock / Inventory** - stock entries, warehouses, item management
- **HR & Payroll** - employees, leave, attendance, payroll
- **Manufacturing** - BOM, work orders, production planning
- **Projects** - tasks, timesheets, project management
- **CRM** - leads, opportunities, and customer interactions

## How to work

1. Use the appropriate MCP tool for each task. Do not try to work around the tools - they are your primary interface to ERPNext.
2. When creating or updating records, confirm the result by reading back the record after the operation.
3. **When you complete a task**, always provide the user with a direct URL link to the relevant ERPNext record so they can verify the result. For example: `<url>/app/sales-order/<name>`.
4. If a tool returns an error, read the error message carefully and adjust your approach - do not retry the exact same call.
5. When listing records, use filters to narrow results rather than fetching everything.

\end{tcblisting}

\clearpage
\subsection{Web Agent Prompt for ServiceNow}

Prompts for other platforms are very similar, with names changed in the first sentence.

\begin{tcblisting}{
    colback=gray!5,
    colframe=black!60,
    title={Prompt for Web Agent for ServiceNow},
    breakable,
    listing only
}
You are a ServiceNow assistant with a Playwright-controlled browser. The browser is pre-configured with authentication headers, so all requests to the ServiceNow instance are already authenticated - you do not need to log in.

## ServiceNow Instance

URL: <url>

When the user asks you to perform a task, navigate to the instance URL above as your starting point unless a more specific URL is provided.

## Capabilities

- **Navigate** to any URL
- **Click** buttons, links, and other elements
- **Fill** forms and input fields
- **Select** dropdown options and checkboxes
- **Screenshot** pages for visual verification
- **Extract** text content and page structure
- **Wait** for elements or network activity
- **Execute JavaScript** in the page context

## How to work

1. When asked to visit a page, use the navigate tool to go there first.
2. Use snapshot/accessibility tools to understand the page structure before interacting with elements.
3. Interact with elements using their accessibility roles and names.
4. After performing actions, verify the result by taking a snapshot or screenshot.
5. Report back what you see and any relevant content from the page.
6. **When you complete a task**, provide the user with a direct URL link to the relevant ServiceNow record or page so they can verify the result.

## Tips

- Always check the page state after navigation or interaction.
- If an element is not found, try taking a snapshot to see what's on the page.
- For complex forms, fill fields one at a time and verify each step.
- Use screenshots when visual context would help the user understand the result.
- ServiceNow uses iframes extensively - you may need to interact with frames.

\end{tcblisting}

\clearpage
\subsection{Terminal Agent Prompt for ServiceNow}

Prompts for other platforms are very similar, with names changed in the first sentence and some platform-specific example changed.
\begin{tcblisting}{
    colback=gray!5,
    colframe=black!60,
    title={Prompt for Terminal Agent for ServiceNow},
    breakable,
    listing only
}
You are a ServiceNow assistant with terminal access. Use bash commands to interact with ServiceNow instances.

## API calls

Instance URL: <url>
$SERVICENOW_EXTRA_HTTP_HEADERS includes all auth headers. Use eval so the flags expand correctly:
```
eval curl -s $SERVICENOW_EXTRA_HTTP_HEADERS \\
  -H '"Content-Type: application/json"' \\
  '<url>/api/now/table/incident?sysparm_limit=5'
```

Filter and sort with `sysparm_query`, select fields with `sysparm_fields`:
```
eval curl -s $SERVICENOW_EXTRA_HTTP_HEADERS \\
  -H '"Content-Type: application/json"' \\
  '<url>/api/now/table/incident?sysparm_query=active=true^ORDERBYDESCsys_created_on' \\
  '&sysparm_fields=number,short_description,state&sysparm_limit=10'
```

POST example:
```
eval curl -s -X POST $SERVICENOW_EXTRA_HTTP_HEADERS \\
  -H '"Content-Type: application/json"' \\
  -d '{"short_description": "example"}' \\
  '<url>/api/now/table/incident'
```

## How to work

- Call ServiceNow APIs with `curl`, referencing $ENV vars for auth.
- Keep commands short. Pipe through `head` to avoid flooding output.
- **When you complete a task**, always provide the user with a direct URL link to the relevant ServiceNow record or page so they can verify the result. For example: `<url>/now/nav/ui/classic/params/target/incident.do?sys_id=<sys_id>`.

\end{tcblisting}

\clearpage
\subsection{API-call Agent Prompt for ServiceNow}

Prompts for other platforms are very similar, with names changed in the first sentence and some platform-specific examples changed.
\begin{tcblisting}{
    colback=gray!5,
    colframe=black!60,
    title={Prompt for API-call Agent for ServiceNow},
    breakable,
    listing only
}
You are a ServiceNow assistant with a single `api_call` tool. Authentication is handled automatically. Use your own knowledge of the ServiceNow REST API to construct the method, path, and JSON body for each call - you do not have access to documentation or a filesystem.

## ServiceNow Instance

URL: <url>

## How to work

1. Use the `api_call` tool to make requests to the ServiceNow REST API. Construct the method, path, and JSON body yourself from your own knowledge of the API - you do not have docs or a filesystem to consult.
2. When creating or updating records, confirm the result by reading back the record after the operation.
3. **When you complete a task**, always provide the user with a direct URL link to the relevant ServiceNow record so they can verify the result. For example: `<url>/now/nav/ui/classic/params/target/incident.do?sys_id=<sys_id>`.
4. If a request returns an error, read the response body carefully and adjust your approach - do not retry the exact same call.
5. When listing records, use query parameters to filter and limit results rather than fetching everything.
- **When the task involves sorting or filtering a list**, always return a URL with explicit `sysparm_query` parameters that reproduce the result. For example:
  `<url>/now/nav/ui/classic/params/target/incident_list.do?sysparm_query=ORDERBYDESCpriority`
\end{tcblisting}

\clearpage

\subsection{Prompt extension for Docs}

\begin{tcblisting}{
    colback=gray!5,
    colframe=black!60,
    title={Prompt Extension for using Docs with ServiceNow},
    listing only
}
## Layout                                                                                     

- **docs/**  - Markdown files of ServiceNow documentation organized by topic hierarchy (e.g. docs/integrate/inbound-rest/concept/c_TableAPI.md). Each file has YAML frontmatter followed by markdown content.

...

## How to work

- **Always consult the docs/ directory first** before making API calls. Look up the relevant endpoint, required parameters, and expected behavior. Browse with `ls`, `find`, `tree`; search with `grep -rl`; read with `cat`, `head`, `wc -l`; or any other terminal tools you find useful.
\end{tcblisting}

\clearpage

\subsection{Prompt extension for Skills}
\begin{tcblisting}{
    colback=gray!5,
    colframe=black!60,
    breakable,
    title={Prompt Extension for using Skills},
    listing only
}
## Layout

- **skills/** - Your persistent memory. Search it before each task; update it after.

...

## Using skills (your memory)

The `skills/` directory is your persistent memory across tasks. It contains reusable procedures, API knowledge, and lessons learned from previous sessions as markdown files. **Always search it before starting a task.** If `skills/` is empty or nothing matches your task, proceed directly.

Useful commands for working with skills:
- **List files**: `ls`, `find`, `tree`
- **Search contents**: `grep`, `grep -rl "<keyword>" skills/`
- **Read files**: `cat`, `head`, `tail`
- **Edit files**: `sed`, `awk`, or rewrite with `cat > skills/path.md << 'EOF'`
- **Create/delete**: `mkdir -p`, `rm`, `mv`

### Reading skills

- If a relevant skill exists, read it and use it as a starting point.
- Check the **Status** field at the top of each skill:
  - `verified` -- confirmed to work; follow with confidence.
  - `unverified` -- use it but verify the result carefully.
- Skills can contain outdated or subtly wrong information -- trust what you observe in the live system over what the skill says.

### Writing skills

When you discover a useful procedure or learn something worth remembering, write it as a skill. Use the following template as a guideline:

```markdown
# <Descriptive Title>

**Status:** unverified

## When to use
<1-2 sentences describing when this skill applies>

## Procedure
<Numbered steps with working commands/API calls>

## Important details
<Field names, parameter values, gotchas>

## Pitfalls
<What NOT to do -- failed approaches and why they fail>
```

Guidelines:

- **Generalize**: write procedures for a *class* of tasks, not one specific instance. Use placeholder values like `PROJECT_NAME`, `USER_ID`, etc. Never hardcode instance URLs -- use `$INSTANCE_URL` or reference the environment variable instead.
- **Include working examples**: paste actual commands and API calls that you have confirmed work.
- **Record failures**: whenever an approach fails, document what you tried and why it didn't work in the Pitfalls section. This prevents repeating the same mistake.
- **New skills start as `unverified`**. Update to `verified` once you have successfully used the procedure on a later task.

### Updating and pruning skills

Skills are a **living knowledge base** that should improve over time:

- **Update, don't duplicate**: if you learn something new about an existing topic, edit the existing file rather than creating a new one.
- **Correct wrong skills**: if a skill doesn't work, fix the procedure and add the failure to the Pitfalls section.

### Organizing skills

- **One file per topic**. Use descriptive filenames (e.g., `create_incident_via_api.md` not `skill_1.md`).
- **Group by knowledge type** using subdirectories (e.g., `procedures/`, `api/`, `troubleshooting/`).

### After completing a task -- reflect

If you learned something genuinely new -- a working procedure, a non-obvious field name, a failed approach worth avoiding -- update or create a skill. Do not write a skill if it would duplicate what is already documented.

\end{tcblisting}

\clearpage

\subsection{Prompts for the Planner and Executor in the Multi-Agent System}

\begin{tcblisting}{
    colback=gray!5,
    colframe=black!60,
    title={Prompt for the Planner in the Planner/Executor Multi-Agent System},
    listing only
}

You are the PLANNER in a two-phase multi-agent system for ServiceNow. You have terminal access to the live instance.

Your job is to research the live instance and produce a detailed, numbered, step-by-step plan for completing the user's task.

## Rules

- You may make read-only API calls (GET requests) to discover available endpoints, field names, or current state.
- Do **NOT** make any state-changing API calls (POST, PUT, PATCH, DELETE).
- Do **NOT** attempt to complete the task yourself.

## Output format

End your response with a clearly labelled plan:

### Plan
1. <step>
2. <step>
...

Include specific details: endpoint paths, field names, parameter values, and any information the executor will need.

\end{tcblisting}

\begin{tcblisting}{
    colback=gray!5,
    colframe=black!60,
    title={Prompt for the Executor in the Planner/Executor Multi-Agent System},
    listing only
}

You are the EXECUTOR in a two-phase multi-agent system for ServiceNow. You have terminal access to the live instance.

A planner has already researched the task and produced a step-by-step plan for you. Your job is to follow the plan and complete the task.

## Rules

- Follow the plan step by step.
- If a step fails or the plan has a mistake, adapt intelligently -- but stay as close to the plan as possible.
- When you complete the task, always provide the user with a direct URL link to the relevant record or page so they can verify the result.

... 

## Plan from the planner

1. **Navigate to the ...

\end{tcblisting}

\clearpage

\subsection{Prompt for the Hybrid Agent}

\begin{tcblisting}{
    colback=gray!5,
    colframe=black!60,
    breakable,
    title={Prompt for the Hybrid Agent},
    listing only
}
You are a ServiceNow assistant with terminal access and a Playwright-controlled browser. Use bash commands for API calls and data processing. Use the browser for tasks that require navigating the ServiceNow web UI, filling forms, or interacting with elements not accessible via API. The browser is pre-authenticated -- you do not need to log in.

## API calls

Instance URL: <url>
$SERVICENOW_EXTRA_HTTP_HEADERS includes all auth headers. Use eval so the flags expand correctly:
```
eval curl -s $SERVICENOW_EXTRA_HTTP_HEADERS \\
-H '"Content-Type: application/json"' \\
'<url>/api/now/table/incident?sysparm_limit=5'
```

Filter and sort with `sysparm_query`, select fields with `sysparm_fields`:
```
eval curl -s $SERVICENOW_EXTRA_HTTP_HEADERS \\
-H '"Content-Type: application/json"' \\
'<url>/api/now/table/incident?sysparm_query=active=true^ORDERBYDESCsys_created_on&' \\
'sysparm_fields=number,short
_description,state&sysparm_limit=10'
```

POST example:
```
eval curl -s -X POST $SERVICENOW_EXTRA_HTTP_HEADERS \\
-H '"Content-Type: application/json"' \\
-d '{"short_description": "example"}' \\
'<url>/api/now/table/incident'
```

## Browser use

When the user asks you to perform a task, navigate to the instance URL above as your starting point unless a more specific URL is provided.

## Capabilities

- **Navigate** to any URL
- **Click** buttons, links, and other elements
- **Fill** forms and input fields
- **Select** dropdown options and checkboxes
- **Screenshot** pages for visual verification
- **Extract** text content and page structure
- **Wait** for elements or network activity
- **Execute JavaScript** in the page context

## How to work

You have two toolsets: a **terminal** (bash commands, curl, scripts) and a **browser** (Playwright). Choose the right tool for the job:

### Terminal (preferred for data operations)
- Call ServiceNow APIs with `curl`, referencing $ENV vars for auth.
- Keep commands short. Pipe through `head` to avoid flooding output.
- Write scripts for repetitive or bulk operations.

### Browser (for UI-specific tasks)
- Use the browser when the task requires navigating to a specific page, filling forms, clicking buttons, or reading UI elements not exposed via API.
- Use snapshot/accessibility tools to understand the page before interacting.
- Interact with elements using their accessibility roles and names.

### General
- **Prefer the terminal for creating, updating, and querying records** -- it is faster and more reliable than the browser for data operations.
- **Use the browser when the API does not support the operation** or when the task explicitly involves the UI.
- You can freely switch between terminal and browser within a task.
- **If an approach fails twice with the same error**, try a fundamentally different strategy -- including switching toolsets.
- **Before finishing a task**, verify your work by querying the live system.
- **When you complete a task**, always provide the user with a direct URL link to the relevant ServiceNow record or page so they can verify the result. For
example: `<url>/now/nav/ui/classic/params/target/incident.do?sys_id=<sys_id>`.
- **When the task involves sorting or filtering a list**, always return a URL with explicit `sysparm_query` parameters that reproduce the result. For example:
`<url>/now/nav/ui/classic/params/target/incident_list.do?sysparm_query=ORDERBYDESCpriority`

## Tips

- Always check the page state after navigation or interaction.
- If an element is not found, try taking a snapshot to see what's on the page.
- For complex forms, fill fields one at a time and verify each step.
- Use screenshots when visual context would help the user understand the result.
- ServiceNow uses iframes extensively -- you may need to interact with frames.

\end{tcblisting}

\clearpage

\section{Error Analysis}\label{sec:error-analysis}

\vspace{-5pt}
\subsection{Case Study: Documentation Leading to a Suboptimal API Strategy}
\label{sec:doc-detrimental-case-study}

We illustrate a case where documentation access was actively detrimental, causing the agent to adopt a more complex API strategy that ultimately failed.
The task requires the agent to order a ``Standard Laptop'' from the ServiceNow Service Catalog with specific configuration options (software requirements, Adobe Acrobat, Adobe Photoshop).

\vspace{-5pt}
\paragraph{Without documentation (score = 1, 7 tool calls).}
The agent discovers the catalog item, retrieves its variables, and attempts to place the order.
After two failed attempts due to shell quoting issues with \texttt{eval}, the agent writes the JSON payload to a file and calls the \texttt{order\_now} endpoint directly:

\begin{lstlisting}[basicstyle=\ttfamily\footnotesize, frame=single, backgroundcolor=\color{gray!5}]
> POST /api/sn_sc/servicecatalog/items/{sys_id}/order_now
  -d @/tmp/order_payload.json
  # payload: {"sysparm_quantity": "1",
  #   "variables": {"Additional_software_...":
  #     "Slack, ...", "acrobat": "true",
  #     "photoshop": "true"}}
-> {"result": {"request_number": "REQ0010001",
    "request_id": "36a3..."}}
\end{lstlisting}
\vspace{-5pt}

The \texttt{order\_now} endpoint accepts the item ID, quantity, and variables in a single call, atomically creating the order with the correct configuration.
The task succeeds.

\vspace{-5pt}
\paragraph{With documentation (score = 0, 19 tool calls).}
The agent begins by searching the documentation for the Service Catalog API.
After 12 tool calls spent on doc retrieval (63\% of total), it finds \texttt{c\_ServiceCatalogAPI.md}, which describes the full catalog API surface including the two-step ordering workflow:
\vspace{-5pt}

\begin{quote}
``POST /sn\_sc/servicecatalog/items/\{sys\_id\}/add\_to\_cart -- Adds the specified item to the cart of the current user.''\\[4pt]
``POST /sn\_sc/servicecatalog/cart/checkout -- Retrieves and processes the checkout for the current cart [\ldots]''
\end{quote}
\vspace{-5pt}
The documentation describes \texttt{add\_to\_cart} followed by \texttt{checkout} as the canonical ordering flow.
Critically, it does not mention the simpler \texttt{order\_now} endpoint that handles everything in one call.
The agent follows the documented approach:

\begin{lstlisting}[basicstyle=\ttfamily\footnotesize, frame=single, backgroundcolor=\color{gray!5}]
# Step 1: Add to cart with variables
> POST /api/sn_sc/servicecatalog/items/{sys_id}
       /add_to_cart
  -d '{"sysparm_quantity": "1",
       "variables": {...}}'
-> {"result": {"cart_id": "27dc...",
    "items": [{"item_name": "Standard Laptop",
               "cart_item_id": "e3dc..."}]}}

# Step 2: Checkout the cart
> POST /api/sn_sc/servicecatalog/cart/checkout
-> {"result": {"request_number": "REQ0010001",
    "request_id": "54ec..."}}
\end{lstlisting}
\vspace{-5pt}
The cart is created and checked out successfully, producing a valid request number.
However, the \texttt{add\_to\_cart} endpoint does not persist the catalog variables (software options) to the resulting request item---it only adds the base item to the cart.
The variables must be set separately via an additional API call that the documentation does not clearly describe.
As a result, the order is placed without the required configuration, and the task fails validation.

\paragraph{Analysis.}
The documentation was \emph{technically correct} but \emph{practically misleading}.
It described a multi-step workflow (\texttt{add\_to\_cart} $\rightarrow$ \texttt{checkout}) that is the general-purpose approach for complex cart operations, but omitted the simpler \texttt{order\_now} endpoint that handles variables atomically.
The agent without documentation discovered \texttt{order\_now} through direct API exploration: a simpler path that happened to be the correct one for this task.

\subsection{Case Study: Documentation Enabling Correct Field Discovery} \label{sec:doc-case-study}

We illustrate the impact of documentation access with a concrete example from the ERPNext evaluation.
The task requires the agent to create a customer record named ``Duck Place'' with several dependencies (user, customer group, tax category), configure it to allow sales invoice creation without sales orders, and add a comment.
The critical subtask is enabling the \texttt{so\_required} field, a non-obvious boolean on the Customer doctype that controls whether sales invoices can be created without a prior sales order.

\paragraph{Without documentation (score = 0).}
The agent successfully creates all prerequisite records and the customer itself in 23 tool calls.
However, when it reaches the sales invoice requirement, it queries the Customer record and observes:

\begin{lstlisting}[basicstyle=\ttfamily\small, frame=single, backgroundcolor=\color{gray!5}]
> GET /api/resource/Customer/Duck Place
  ?fields=["so_required","dn_required"]
-> so_required : 0
   dn_required : 0
\end{lstlisting}

The agent sees \texttt{so\_required = 0} but does not understand that this field must be set to \texttt{1} to \emph{allow} invoice creation without a sales order.
The field name is counterintuitive: \texttt{so\_required} controls whether a sales order is \emph{required}, but the task asks to \emph{allow} invoices \emph{without} one.
The agent leaves the value at its default (0) and moves on, resulting in a failed evaluation.

\paragraph{With documentation (score = 1).}
The agent follows a similar trajectory but, after creating the customer, reads the ERPNext Customer documentation:

\begin{lstlisting}[basicstyle=\ttfamily\small, frame=single, backgroundcolor=\color{gray!5}]
> cat docs/erpnext/user/manual/en/customer.md
\end{lstlisting}

The documentation contains the following passage:

\begin{quote}
\small
``If the `Delivery Note Required' or `Sales Order Required' option is configured as `Yes' in Selling Settings, it can be overridden for a particular customer by enabling the `Allow Sales Invoice Creation Without Sales Order' [\ldots]''
\end{quote}

This tells the agent two things: (1) the feature exists as a per-customer override, and (2) it is related to sales order requirements.
The agent then queries the Customer DocType schema to find the exact field name:

\begin{lstlisting}[basicstyle=\ttfamily\small, frame=single, backgroundcolor=\color{gray!5}]
> GET /api/resource/DocType/Customer
  | grep "so_required\|dn_required"
-> so_required | Allow Sales Invoice Creation Without
                 Sales Order
   dn_required | Allow Sales Invoice Creation Without
                 Delivery Note
\end{lstlisting}

With this understanding, the agent correctly sets \texttt{so\_required = 1}:

\begin{lstlisting}[basicstyle=\ttfamily\small, frame=single, backgroundcolor=\color{gray!5}]
> PUT /api/resource/Customer/Duck Place
  -d '{"so_required": 1}'
-> 200 OK
\end{lstlisting}

The task completes successfully in 27 tool calls (5 spent on documentation).

\paragraph{Analysis.}
The documentation did not provide the API call or the field name directly---the agent still needed to query the schema.
Rather, it provided \emph{conceptual awareness} that the feature existed and was configurable per customer, prompting the agent to investigate the correct field.
Without this nudge, the agent saw \texttt{so\_required = 0}, interpreted it as already correct, and moved on.
The pattern of documentation providing conceptual scaffolding rather than exact commands is characteristic of cases where documentation is most helpful.

\subsection{Case Study: Skills Reducing Exploration Overhead}
\label{sec:skills-case-study}

We illustrate the efficiency benefit of persistent memory with a ServiceNow change request creation task (\texttt{create-change-request.957}).
The task requires setting 12 fields, including several whose valid values must be discovered at runtime: \texttt{risk} (``Moderate''), \texttt{impact} (``2 - Medium''), \texttt{category} (``Network''), and \texttt{close\_code} (``Successful with issues'').
Both agents succeed, but the agent with skills completes the task in 5 tool calls at \$0.34, while the baseline requires 8 tool calls at \$0.51---a 33\% reduction in cost.

\paragraph{Without skills (8 calls, \$0.51).}
The agent queries the \texttt{sys\_choice} table for valid field values, but impact choices are missing from the \texttt{change\_request} table:

\begin{lstlisting}[basicstyle=\ttfamily\small, frame=single, backgroundcolor=\color{gray!5}]
> GET /api/now/table/sys_choice
  ?sysparm_query=name=change_request
    ^element=impact
-> (no results)
\end{lstlisting}

After a second failed attempt, the agent discovers that impact choices are inherited from the parent \texttt{task} table:

\begin{lstlisting}[basicstyle=\ttfamily\small, frame=single, backgroundcolor=\color{gray!5}]
> GET /api/now/table/sys_choice
  ?sysparm_query=name=task^element=impact
-> "1" = High, "2" = Medium, "3" = Low
\end{lstlisting}

The agent then encounters a shell escaping error when attempting an inline JSON POST, requiring an additional call to write the payload to a temporary file before succeeding.

\paragraph{With skills (5 calls, \$0.34).}
The agent reads \texttt{create\_change\_request.md}, a skill file accumulated from prior tasks:

\begin{lstlisting}[basicstyle=\ttfamily\small, frame=single, backgroundcolor=\color{gray!5}]
> grep -rl "change" skills/
> cat skills/procedures/create_change_request.md
\end{lstlisting}

The skill contains the field-to-value mappings the baseline agent had to discover:

\begin{quote}
\small
``Impact: \texttt{"1"}=High, \texttt{"2"}=Medium, \texttt{"3"}=Low (inherited from task table, not change\_request). Risk: \texttt{"1"}=Very High, \texttt{"2"}=High, \texttt{"3"}=Moderate, \texttt{"4"}=Low. Do NOT inline multiline JSON in eval curl---always write to a file and use \texttt{-d '@/tmp/file.json'}.''
\end{quote}

The agent looks up only the instance-specific assignment group \texttt{sys\_id} (not in the skill), writes the payload to a file, and posts it on the first attempt.

\paragraph{Analysis.}
The three calls saved correspond directly to knowledge encoded in the skill: the impact field's inheritance from the \texttt{task} table (two failed discovery calls avoided) and the file-based payload pattern (one failed inline attempt avoided).
Skills amortize the cost of API exploration across tasks, converting runtime discovery into a single file read.

\subsection{Case Study: Planner-Executor Resolving Ambiguous Field Semantics}
\label{sec:mas-case-study}

We illustrate the benefit of multi-agent planning with the same ERPNext task used in Section~\ref{sec:doc-case-study}: creating a customer record with sales invoice configuration.
The critical subtask is setting \texttt{so\_required = 1} to allow sales invoices without a sales order: a field whose name suggests the opposite of its effect.

\paragraph{Single agent (score = 0).}
The agent creates all prerequisite records and the customer in 18 tool calls.
When it reaches the sales invoice configuration, it queries the Customer record, sees \texttt{so\_required = 0}, interprets this as ``sales orders are not required'' (the desired state), and moves on without modification.
The task fails because the field's semantics are inverted: \texttt{so\_required = 1} means ``allow invoices without sales orders.''

\paragraph{Planner-executor (score = 1).}
The planner spends 22 tool calls exploring the system before producing an execution plan.
Critically, it queries the Customer DocType schema to understand the available fields:

\begin{lstlisting}[basicstyle=\ttfamily\small, frame=single, backgroundcolor=\color{gray!5}]
> GET /api/resource/DocType/Customer
  | grep "so_required"
-> so_required | Allow Sales Invoice Creation Without
                 Sales Order
\end{lstlisting}

By reading the field's human-readable label rather than just its current value, the planner correctly maps the task requirement to \texttt{so\_required = 1}.
It produces a structured plan that includes this step explicitly:

\begin{lstlisting}[basicstyle=\ttfamily\small, frame=single, backgroundcolor=\color{gray!5}]
Step 4: Configure sales invoice settings
  PUT /api/resource/Customer/Duck Place
  -d '{"so_required": 1}'
  # so_required=1 means "Allow Sales Invoice
  # Creation Without Sales Order"
\end{lstlisting}

The executor follows the plan in 8 tool calls, setting the field correctly without needing to reason about its semantics independently.

\paragraph{Analysis.}
Both agents encountered the same ambiguity, but the planner's exploratory phase (querying the DocType schema rather than the record's current value) happened to expose the field's label, which disambiguated its meaning.
The planner's output then acted as a form of structured documentation for the executor, encoding the correct interpretation so that execution became more straightforward.
This example suggests that the planner-executor design may be particularly helpful when tasks involve fields or parameters whose programmatic names do not obviously reflect their semantic intent, and where schema exploration can help resolve such ambiguities before execution begins.

\end{document}